\documentclass[fleqn,usenatbib]{mnras}

\usepackage{newtxtext,newtxmath}
\usepackage[T1]{fontenc}
\usepackage{ae,aecompl}

\usepackage{graphicx}
\usepackage{amsmath,bm,mathtools}
\usepackage{booktabs}
\usepackage{array}
\usepackage{siunitx}
\usepackage{xcolor}

\newcommand{\dd}{\mathrm{d}}
\newcommand{\ii}{\mathrm{i}}
\newcommand{\grad}{\boldsymbol{\nabla}}
\newcommand{\divg}{\boldsymbol{\nabla}\!\cdot}
\newcommand{\ve}[1]{\boldsymbol{#1}}

\newcommand{\Om}{\Omega}
\newcommand{\sig}{\sigma}
\newcommand{\xin}{\boldsymbol{\xi}}
\newcommand{\ue}{\boldsymbol{u}}
\newcommand{\geff}{\boldsymbol{g}_{\rm eff}}

\newcommand{\ajb}[1]{\textcolor{green}{#1}}

\title[2D FEM oscillations of rotating polytropes]
{Finite-Element Asteroseismology I: Two-Dimensional 
Adiabatic Oscillations of Rapidly Rotating $N=3$ Polytropes}

\author[Guo \& Barker]{Zhao Guo$^{1}$ and Adrian J. Barker$^{1}$\\
$^{1}$Department of Applied Mathematics, School of Mathematics, University of Leeds, Leeds LS2 9JT, UK\\
}

\date{Accepted Nov 6 666. Received Nov 6 666; in original form Nov 6 666}
\pubyear{2026}

\begin{document}
\label{firstpage}
\pagerange{\pageref{firstpage}--\pageref{lastpage}}
\maketitle

\begin{abstract} 
We develop a finite-element formulation for linear adiabatic oscillations of axisymmetric uniformly rotating, centrifugally distorted, self-gravitating $N=3$ polytropes. Equilibrium models are constructed using self-consistent-field iteration and then remeshed on a surface-fitted oblate meridional domain. The quadratic eigenvalue problem describing oscillations for each azimuthal wavenumber $m$ is solved using
a conforming two-domain mesh. The
density-weighted displacement,
$\boldsymbol{u}=\rho^{1/2}\boldsymbol{\xi}$, is defined in the stellar interior, whereas the gravitational-potential perturbation $\delta\Phi$ is solved both inside the star and in an exterior vacuum. The coupled displacement--potential problem is assembled and solved within FEniCSx
as a polynomial eigenvalue problem.

Without rotation, the axisymmetric 
modes agree closely with an independent one-dimensional calculation. For finite rotation rates, several clean axisymmetric 
$p$-mode branches of the spheroidal star successfully reproduce the frequencies tabulated by Reese (2006). The relative frequency differences induced by rotation are smaller than  $3\times10^{-4}$ for the low-order mode branches up to $\sim 60$\% of the critical rotation rate. These tests  validate both the centrifugally distorted equilibrium and the finite-element implementation.

We develop a method to track eigenfunctions continuously as a function of rotation and decompose their angular structure to quantify rotational mode mixing and progressive coupling between different spherical harmonic components. This reveals continuous deformation of individual mode branches from nearly single-$\ell$ structures to strongly mixed states at rapid rotation.

This work is intended as a step toward finite-element multidimensional asteroseismology frameworks for modelling realistic rotationally and tidally distorted stellar configurations.

\end{abstract}


\section{Introduction}
\label{sec:intro}

Rapid rotation renders stellar oscillations intrinsically multidimensional: the centrifugal acceleration deforms the equilibrium structures away from spherical symmetry, while the Coriolis acceleration couples motions with different
angular characters \citep{Maeder2009}. For sufficiently slow rotation rates, these effects can be accurately treated as small perturbations to the oscillations of a spherical
stellar model \citep[e.g.][]{Saio1981,GoughThompson1990,
DziembowskiGoode1992,Soufi1998,Suzrez2006,Guo2024}. However, as the rotation rate is increased, higher-order rotational effects, including centrifugal deformation, become
increasingly important, and perturbative treatments become progressively inaccurate \citep[e.g.][]{Ballot2010}. Depending on the mode and stellar model, perturbative approaches can fail to accurately predict oscillations for rotation rates of order $15\%$ of the critical rotation rate,
motivating fully non-perturbative calculations of rapidly rotating configurations. Such approaches have been developed in several related
fields, including asteroseismology, neutron star oscillations, and rotating fluid dynamics.

An early line of development used direct multidimensional
(low-order finite-difference) discretizations of the oscillation equations. Two-dimensional
hydrodynamical calculations of stellar pulsations were performed by \citet{Deupree1977,Deupree1980}, while pioneering direct calculations of the linear adiabatic eigenvalue problem in rapidly rotating stars were presented by
\citet{Clement1981,Clement1984,Clement1986,Clement1989,Clement1998}. Related two-dimensional formulations for rapidly rotating Newtonian stellar models, including perturbed self-gravity, were introduced by \citet{IpserLindblom1989,IpserLindblom1990}. A parallel development took place in the neutron star community, where rapid rotation motivated non-perturbative calculations of $f$-, $r$- and inertial modes in Newtonian and relativistic models
\citep[e.g.][]{YoshidaEriguchi1995,YoshidaEriguchi1997,
YoshidaEriguchi1999,YoshidaEriguchi2001,Karino2000,Karino2001,
Yoshida2002,Yoshida2005}. Later work also used multidimensional
time evolution and relativistic perturbation methods
\citep[e.g.][]{Font2001,GaertigKokkotas2008,Passamonti2009,
Yoshida2012,Doneva2013,KrugerKokkotas2020}.

A separate line of research led to the development of the modern non-perturbative oscillation codes used in
asteroseismology. \citet{Espinosa2004} calculated oscillations in axisymmetrically distorted stellar models, while  \citet{Lignieres2006} and \citet{Reese2006} treated centrifugally
distorted $N=3$ polytropes with a two-dimensional spectral method. The approach was subsequently applied to more realistic and differentially rotating models \citep{Reese2009}, and to inertial--gravity modes \citep{Ballot2010}. The ACOR code of \citet{Ouazzani2012} provided an alternative non-perturbative treatment, which was later extended to more realistic rotating stellar structures \citep{Ouazzani2015}. Other two-dimensional calculations include the NRO code of \citet{LovekinDeupree2008}, while more recent TOP calculations have been performed on fully two-dimensional ESTER (Evolution STEllaire en Rotation) stellar models \citep{Rieutord2016,Reese2021}.

Closely related problems have also been studied in the rotating-fluid and geophysical-fluid communities. Inertial, acoustic, and hydromagnetic
modes in spheroidal and triaxial ellipsoids have been calculated using
global polynomial and spectral methods
\citep[e.g.][]{L1989a,L1989b,Vantieghem2014,BBO2016,BackusRieutord2017,VidalCebron2017,
Vidal2020,VidalCebron2020,VidalCebron2021,VidalDynamo2021}. These have been applied to tidally forced problems in ellipsoids or oblate spheroidal polytropes \citep[e.g.][]{BBO2016,Barker2016,Dewberry2023}. These studies are particularly relevant because they address global eigenmodes in strongly non-spherical rotating domains and, in several cases, use finite-element or spectral-element calculations as independent numerical benchmarks. In particular \citet{Barker2016} used spectral-element methods to solve nonlinear time-dependent tidal flows in a homogeneous ellipsoidal model of a rotating and tidally deformed star or planet.

Finite-element methods (FEM) have also been applied to a variety of oscillation and wave-eigenvalue problems.
An early application to stellar pulsations was given by
\citet{FiedlerAndrew1986}, who used FEM for
non-radial oscillations of spherical stellar models. For rotating-fluid
problems, \citet{SchulkesCuvelier1991} formulated the normal-mode problem
of a rotating fluid with a free boundary as a quadratic
eigenvalue problem using FEM. Finite-element eigenmode calculations have also been
used successfully in strongly non-spherical geometries, for example for
axisymmetric resonators at fixed azimuthal order \citep{Oxborrow2007},
for acoustic modes and rotational splittings in rotating spheroids
\citep[e.g.][]{Su2020}, and for acoustic modes of rapidly rotating ellipsoids
subject to centrifugal gravity \citep[e.g.][]{VidalCebron2021}. More recently,
\citet{Faucher2025} developed a three-dimensional element-based
discretization for time-harmonic solar oscillations. These studies
demonstrate the suitability of finite-element (or spectral-element) approaches for oscillation
problems involving non-spherical
geometries, and motivate their
application here to the global eigenmodes of centrifugally deformed, self-gravitating stellar models.

Despite this extensive literature, FEM have been used much less widely for global stellar oscillation eigenvalue problems than finite-difference or spectral methods. In this work, we develop a two-dimensional FEM formulation for
the linear adiabatic oscillations of uniformly rotating, centrifugally distorted $N=3$ polytropes. The equilibrium models are constructed with a
self-consistent-field method, and the oscillation equations are solved for each azimuthal order $m$ using quadratic finite elements in FEniCSx \citep{alnaes2014ufl,baratta2025}. The displacement is solved in the oblate stellar domain, while Eulerian perturbations of the gravitational potential $\delta\Phi$ are also solved in a conforming exterior vacuum domain. The resulting mixed problem is written as a quadratic polynomial eigenvalue problem and solved with the linear algebra packages PETSc/SLEPc \citep{slepc2005,Petsc2026}. We validate the formulation against independent spherical
calculations and against the non-perturbative frequencies of \citet{Reese2006a},



The finite-element method (FEM) provides a complementary approach to spectral formulations. Its principal advantage is geometrical flexibility: an unstructured or suitably adapted mesh can conform directly to a distorted stellar surface, without requiring the equilibrium shape to be represented by a small set of spherical harmonics. This becomes particularly useful when the equilibrium is not axisymmetric. For example, in a tidally distorted star the background structure generally depends on both latitude and longitude, so different $(\ell,m)$ spherical harmonic components of the oscillation problem are coupled and the usual separation in azimuthal order is lost. A finite-element
formulation can instead represent such geometries directly in physical space, making it attractive for calculations of rapidly rotating and tidally distorted stars.


To facilitate reproducibility and wider community use, we will make the FEniCSx scripts developed in this work openly available, which provides an open-source alternative to established multidimensional oscillation codes such
as TOP and ACOR. 

The paper is organized as follows. Section~\ref{sec:fem} presents the eigenvalue problem (EVP) in FEM.
Section~\ref{sec:scf} describes the equilibrium rotating polytrope model generated using the Self-Consistent Field (SCF) method \citep{Ostriker1968,Hachisu1986}.
Section \ref{sec:equations} presents the adiabatic oscillation equations in both their strong and weak forms. The classical 1D oscillation equations are reviewed in section 5, and section 6 presents the validation tests and rotating-mode comparisons for our 2D oscillations. Section~\ref{sec:eigenfunctions} 
discusses mode eigenfunction and classification, and we present a method to track modes as a function of rotation. 
The principal limitations and some future extensions are discussed in Sections~\ref{sec:limitations} and \ref{sec:prospects}, respectively, followed by the conclusions in Section~\ref{sec:conclusions}. In the Appendix, we also show additional calculations computed that adopt the Cowling approximation, and provide further details of the FEM discretization.

  
\section{Finite-element formulation and FEniCSx implementation}
\label{sec:fem}

The oscillation problem in a rapidly rotating star is  intrinsically multidimensional because the centrifugal acceleration deforms both the equilibrium structure and the stellar surface from spherical symmetry. In contrast with a spherical harmonic expansion that is truncated at a finite angular degree, the present calculation solves the oscillations directly on a two-dimensional meridional 
domain, which is appropriate for describing modes in an axisymmetric stellar model. The finite-element method is particularly convenient for this purpose because the computational mesh can conform to an arbitrary oblate boundary, while the weak variational formulation naturally accommodates spatially varying equilibrium coefficients and regularity conditions on the rotation axis.

We use cylindrical polar coordinates $(s,\phi,z)$, with the rotation axis parallel to $\ve{e}_z$, and solve only on the meridional half-plane $s\geq0$. For a prescribed azimuthal order $m$, every perturbed quantity has the dependence
\begin{equation}
    \delta X(s,z,\phi,t)
    = \widehat{\delta X}(s,z)
      \exp(\ii m\phi-\ii\sig t),
\end{equation}
where $\sig$ is the mode frequency in the uniformly rotating frame. The corresponding inertial-frame frequency is $ \omega_{\rm in}=\sig+m\Om $. This convention is used throughout this work and the numerical implementation.

The finite-element state vector consists of a modified displacement vector field and the 
gravitational potential,
\begin{equation}
    X =
    \left(
    \boldsymbol{u},\,\delta\Phi
    \right),
\end{equation}
where $\boldsymbol{u}=\rho^{1/2}\xin=(u_+,u_-,u_z)$ denotes the density-weighted displacement using a real helical representation (see Eq.~\ref{eq:helical} later). Note that $\boldsymbol{u}$ does not represent the Eulerian velocity field in this work. Also note that the density and enthalpy perturbations are not independent variables: the continuity equation and adiabatic relation determine them directly from $\boldsymbol{\xi}$. We use quadratic ($P2$) Lagrange elements for both $\boldsymbol{u}$ and $\delta\Phi$. This choice was made for simplicity and computational efficiency. Future calculations should explore higher order elements to improve accuracy for a given number of elements.


The displacement degrees of freedom are only considered in the stellar subdomain. The values exclusively in the exterior vacuum are constrained to be zero, while the stellar boundary remains free. In contrast,
$\delta\Phi$ is a single continuous finite-element field across the combined stellar and vacuum domains.

The background enthalpy profile is represented by a quadratic finite-element field. Near the artificial star-vacuum interface, however, quadratic interpolation of $\rho$ can produce small negative undershoots between otherwise non-negative nodal values, making quantities involving $\rho^{1/2}$ ill defined. To avoid this, $\rho$ and $\rho^{1/2}$ are represented separately using continuous linear (P1) elements throughout the computational domain. This lower-order treatment is applied globally, but is primarily needed to maintain a non-negative interpolation near the stellar boundary.

\section{Uniformly rotating $N=3$ polytropes from the SCF method}
\label{sec:scf}


\begin{figure*}
\centering
\includegraphics[height=0.68\textheight]{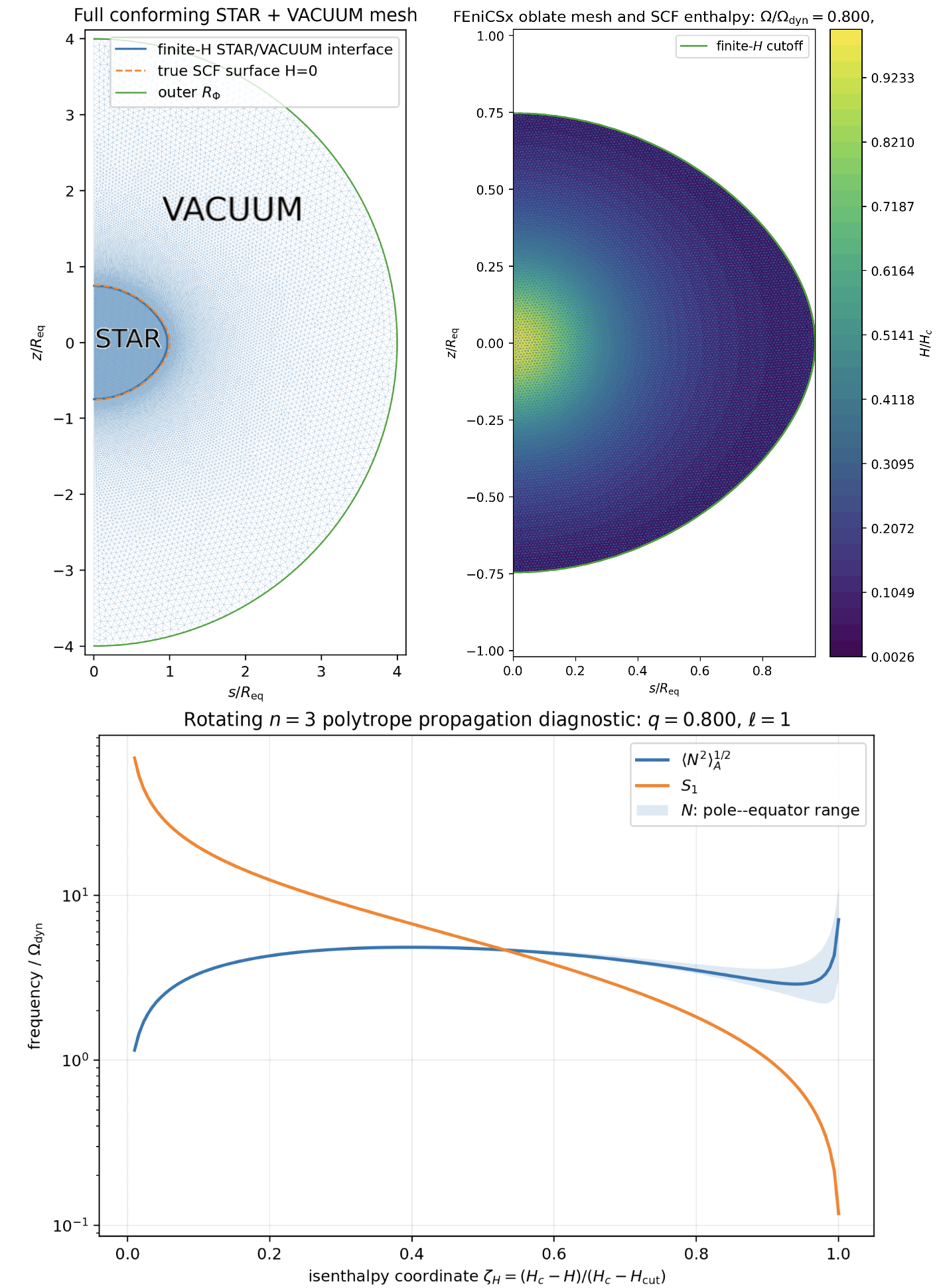}
\caption{\textbf{Top right:} equilibrium enthalpy distributions, normalized by the central enthalpy, 
for the uniformly rotating $N=3$ polytrope models with $q=\Omega/\Omega_{\rm dyn}=0.8$. 
\textbf{Top left:} Two-domain finite-element meshes for the full-gravity oscillation
problem. The inner region is the centrifugally distorted stellar domain bounded by the finite-enthalpy contour $H=H_{\rm cut}$, while the exterior vacuum extends to the circular boundary $R_{\Phi}=4R_{\rm eq}$. The mesh has characteristic resolution $h_{\star}=0.01R_{\rm eq}$ near the star and is smoothly coarsened to $h_{\rm vac}=0.10R_{\rm eq}$ in the far vacuum. \textbf{Lower:} Brunt--V\"ais\"al\"a frequency $N$ and Lamb frequency
$S_{\ell=1}$ as functions of the iso-enthalpy coordinate $\zeta_H$ for
the same polytropic model. The subscript $A$ of Brunt frequency means surface-area average over an isenthalpy shell. The Lamb frequency is defined by
$S_\ell=\sqrt{\ell(\ell+1)}\,c_s/R_{\rm vol}$, where $c_s$ is the adiabatic sound speed and $R_{\rm vol}$ is the volume-equivalent radius of the corresponding iso-enthalpy surface.}

\label{fig:scf_background}
\end{figure*}


We construct uniformly rotating barotropic equilibria using a self-consistent-field (SCF) iteration method \citep{Ostriker1968,Hachisu1986}. The adopted equation of state is
\begin{equation}
    P=K\rho^{1+1/N},
    \qquad \text{with} \qquad N=3,
\end{equation}
for which the specific enthalpy is
\begin{equation}
    H(\rho)
    =\int_0^P\frac{\dd P'}{\rho(P')}
    =(N+1)K\rho^{1/N}
    =4K\rho^{1/3}.
\end{equation}
Hence,
\begin{equation}
    \rho=\left(\frac{H}{4K}\right)^3 \qquad \text{where} \qquad
    \quad (H>0).
    \label{eq:rho_from_H}
\end{equation}
with $\rho=0$ where $H\leq0$. An $N=3$ polytrope provides a simple idealized model of a strongly centrally condensed massive star.


For uniform rotation, hydrostatic equilibrium may be written as the Bernoulli relation
\begin{equation}
    H+\Phi-\frac{1}{2}\Om^2 s^2=C,
    \label{eq:bernoulli_scf}
\end{equation}
where $\Om$ is the angular velocity, and $\Phi$ is the gravitational potential, satisfying
\begin{equation}
    \nabla^2\Phi=4\pi G\rho.
    \label{eq:poisson}
\end{equation}
The SCF iteration alternates between solving Eq.~\eqref{eq:poisson} 
for the gravitational potential generated by the current mass density distribution and updating the enthalpy using Eq.~\eqref{eq:bernoulli_scf}. The density is then recomputed from Eq.~\eqref{eq:rho_from_H}. At each SCF iteration, $C$ is chosen so that the central enthalpy remains
fixed, giving $C=H_c+\Phi_c$ because the centrifugal term vanishes at the
centre. We hold $K$ and $H_c$ fixed during the iteration and subsequently
apply the $N=3$ homology rescaling to enforce the target mass\footnote{With $\lambda_s=M_{\rm target}/M_{\rm SCF}$, we can scale the equilibrium variables as: $r'=\lambda_s r$, $\rho'=\lambda^2_s\rho$, $P'=\lambda^2_s P$, $\Omega'=\lambda^{-1}_s\Omega$, and $K'=\lambda^{2/3}_s K$}. 

To improve numerical stability, under-relaxation is applied to the density update, such that
\begin{equation}
    \rho^{(k+1)}
    =(1-\lambda)\rho^{(k)}
    +\lambda\rho_{\rm new}^{(k+1)},
\end{equation}
where the relaxation parameter $\lambda$, with the default value of $0.35$, is reduced automatically for rapidly rotating models. Convergence is monitored using the relative $L_2$ norm of
the density change, i.e. the volume-weighted root-mean-square difference
between successive density iterates normalized by the density norm, together with integrated and maximum density changes and the changes in the
equatorial and polar radii.

We parameterize the rotation rate of the converged model by
\begin{equation}
    q\equiv
    \frac{\Om}{\sqrt{GM/R_{\rm eq}^3}}=\frac{\Om}{\Omega_{\rm dyn}},
    \label{eq:qdef}
\end{equation}
where $R_{\rm eq}$ is the equatorial radius and $\Omega_{\rm dyn}\equiv\sqrt{GM/R_{\rm eq}^3}$ is the dynamical frequency. Because $R_{\rm eq}$ itself varies during the SCF iteration, the physical angular velocity is adjusted by a safeguarded outer iteration until the requested $q$ is reached. Rapidly rotating models are constructed by continuation using a sequence of intermediate $q$ values, with smaller continuation steps and stronger under-relaxation parameters as the mass-shedding limit ($q\to 1$) is approached.
For uniformly rotating $N=3$ polytropes, the mass-shedding limit is reached before the classical secular or dynamical bar-mode instability \citep{James1964}, so the models considered here up to $q\lesssim1$ remain on the corresponding stable equilibrium branch.

After convergence, the equilibrium is normalized so that
\begin{equation}
    G=M=R_{\rm eq}=1.
\end{equation}
The normalized equilibrium model stores the two-dimensional thermodynamic structure on
the meridional cylindrical coordinates $(s,z)$, including the enthalpy,
density, pressure, and sound speed, together with the stellar surface
radius $R(\theta)$ as a function of colatitude $\theta$. The degree of
centrifugal deformation is characterized by the flattening parameter 
\begin{equation}
    f=1-\frac{R_p}{R_{\rm eq}},
\end{equation}
where $R_p$ is the polar radius, is a useful diagnostic that we compute, as are the scalar virial and hydrostatic residuals. For $q=0$, the numerical equilibrium is also 
compared directly with the one-dimensional Lane--Emden $N=3$ solution. 


Figure~1 shows the structure of a rapidly rotating $N=3$ polytrope with a rotation rate $q=0.8$. The upper right panel shows the enthalpy distribution in the meridional plane, while the upper left panel shows the finite-element mesh used in FEniCSx for the stellar interior and the exterior vacuum domain. The lower panel shows the propagation diagram within the star, displaying the Brunt--V\"ais\"al\"a frequency $N$ and the $\ell=1$ Lamb frequency $S_{\ell=1} \propto c_s$ as functions of the iso-enthalpy coordinate $\zeta_H$. 

\subsection{Two-domain stellar and vacuum mesh}

The full-gravity calculation (meaning that we incorporate the effects of self-gravity on both the equilibrium stellar structure and the perturbations) requires the gravitational potential perturbation to be computed outside the stellar surface. We therefore construct a conforming two-domain mesh in the meridional half-plane $s\ge0$, defined by
\begin{equation}
 {\cal D}={\cal D}_{\star}\cup{\cal D}_{\rm vac}.
\end{equation}
The stellar domain ${\cal D}_{\star}$ is bounded by the finite-enthalpy isosurface
\begin{equation}
 H(s,z)=H_{\rm cut},
\end{equation}
where $H_{\rm cut}/H_{\rm c}$ is chosen to be the value corresponding to $r/R=0.99$ in the non-rotating $N=3$ Lane--Emden model, which has a non-zero (and positive) associated density. The exterior domain ${\cal D}_{\rm vac}$ occupies the region between this oblate interface and a circular (i.e., meridional slice of a spherical boundary) outer boundary at
\begin{equation}
 r=R_{\Phi},
 \qquad R_{\Phi}=4R_{\rm eq}.
\end{equation}

The adopted outer boundary is larger than the
$R_{\Phi}=2R_{\rm eq}$ commonly used in spectral calculations such as TOP and ACOR.  In these methods, the angular dependence of the exterior potential is represented spectrally, allowing each spherical harmonic
component to be matched directly to its analytic vacuum solution, $\delta\Phi_{\ell}\propto r^{-(\ell+1)}$.  In the present FEM formulation, by contrast, $\delta\Phi$ is represented directly on the two-dimensional exterior domain and a simple homogeneous Dirichlet condition,
$\delta\Phi=0$, is imposed at the outer boundary.  Since this condition only
approximates the asymptotic decay at infinity, we place the boundary farther
from the star so that the gravitational potential perturbation has decayed
substantially before reaching it.  The choice $R_{\Phi}=4R_{\rm eq}$ therefore
provides a conservative and numerically straightforward approximation to the
infinite vacuum domain while avoiding the non-local spherical harmonic.

The two subdomains are generated simultaneously using Gmsh
\citep{geuzaine2009gmsh} and share the same discretized stellar interface, so that interpolation is not required to couple the interior and exterior gravitational potential solutions. The stellar interface is first constructed
as a spline through points on the numerically determined finite-enthalpy SCF contour, $r_{\rm cut}(\theta)$. The mesh is subsequently promoted from first-order  to quadratic order, so that the interface is represented in DOLFINx by piecewise-quadratic curved finite-element edges. The boundary is then a quadratic approximation to the SCF isenthalpy contour rather than to an assumed analytic oblate spheroid.

The gravitational potential perturbation satisfies
\begin{equation}
 \nabla^2\delta\Phi=4\pi G\,\delta\rho
 \qquad\hbox{in }{\cal D}_{\star},
 \label{eq:poisson_star}
\end{equation}
and
\begin{equation}
 \nabla^2\delta\Phi=0
 \qquad\hbox{in }{\cal D}_{\rm vac}.
 \label{eq:poisson_vac}
\end{equation}

The computational stellar boundary is a finite-density
surface, so it is convenient to interpret the perturbed Poisson equation in a distributional form on the combined domain via \eqref{eq:poisson_star} 
and \eqref{eq:poisson_vac}. Since $\rho^{1/2}\boldsymbol{u}$ is taken to have zero value in the exterior vacuum, the gravitational potential in the entire domain satisfies
\begin{equation}
 \nabla^2\delta\Phi
 +4\pi G\nabla\cdot
 \left(
 \rho^{1/2}\boldsymbol{u}
 \right)
 =0 ,
 \label{eq:poisson_distribution}
\end{equation}
where the divergence operator is understood to apply within
${\cal D}_{\star}\cup{\cal D}_{\rm vac}$. In the weak formulation, this automatically includes the interface contribution associated with the displacement of a finite-density boundary. Consequently, we do not need to add a separate surface mass source by hand.


A characteristic element size $h_{\star}=0.01R_{\rm eq}$ is adopted on, and near, the stellar boundary. In the exterior potential domain, the prescribed mesh size is smoothly increased linearly with distance from the stellar interface, using Gmsh's distance-based threshold field, until it reaches the characteristic far-field size $h_{\rm vac}=0.10R_{\rm eq}$. This is sufficient because
$\delta\Phi$ varies smoothly in the vacuum region, and contains no short-scale structure there. The fine resolution is retained across the common star--vacuum interface, since this is where accurate coupling between the displacement and gravitational potential is most important.

For the $N=3$ polytrope, all thermodynamic coefficients needed by the oscillation operator are reconstructed from the interpolated enthalpy field. This preserves the internal hydrostatic consistency of the background stellar model.  

In the normalization used below,
\begin{equation}
\begin{aligned}
    \geff &= \grad H,
    \qquad
    \grad\ln\rho = 3\frac{\grad H}{H},
    \\
    c_s^2 &= \frac{\Gamma_1 H}{4},
    \qquad
    N^2 =
    \left(
    3-\frac{4}{\Gamma_1}
    \right)
    \frac{|\nabla H|^2}{H},
\end{aligned}
\label{eq:n3_background_identities}
\end{equation}
with $\Gamma_1=5/3$ (an ideal monatomic gas). Since the
$N=3$ polytropic equilibrium has $P\propto\rho^{4/3}$, this choice gives $\Gamma_1>4/3$ and therefore it is a stably stratified configuration with
$N^2>0$ (Fig. 1, lower panel), rather than a neutrally or convectively unstable model. In this
sense, the model is most representative of radiative stellar regions, such as the envelopes of intermediate- and high-mass main-sequence stars. 

\section{Linear adiabatic oscillations}
\label{sec:equations}

\subsection{Governing equations in the rotating frame}

Let $\boldsymbol{\xi}$ be the Lagrangian displacement and adopt
the dependence $\propto  \exp(i m\phi-i\sigma t)$, where $\sigma$ is defined in the uniformly rotating frame. The
linearized momentum equation, retaining the Eulerian perturbation of the gravitational potential, can be written
\begin{equation}
 -\sigma^2\rho\boldsymbol{\xi}
 -2i\sigma\rho\boldsymbol{\Omega}\times\boldsymbol{\xi}
 =
 -\nabla\delta P
 +\delta\rho\,\boldsymbol{g}_{\rm eff}
 -\rho\nabla\delta\Phi ,
 \label{eq:momentum_fullgravity}
\end{equation}
where
\begin{equation}
 \boldsymbol{g}_{\rm eff}=\frac{\nabla P}{\rho}=\nabla H ,
\end{equation}
and $\delta P$, $\delta \rho$ are Eulerian perturbations of the pressure and density.
Mass conservation and the assumption of adiabatic perturbations give
\begin{align}
 \delta\rho &=
 -\nabla\cdot(\rho\boldsymbol{\xi}),
 \\
 \delta P &=
 -\Gamma_1P\nabla\cdot\boldsymbol{\xi}
 -\boldsymbol{\xi}\cdot\nabla P .
\end{align}

A key step in our numerical method is to introduce the density-weighted displacement
\begin{equation}
    \ue=\rho^{1/2}\xin.
    \label{eq:u_def}
\end{equation}
For any vector field $\ue$, we can then define
\begin{align}
    D(\ue)&\equiv\divg\ue,\\
    Q(\ue)&\equiv D(\ue)
       -\frac{1}{2}\ue\cdot\grad\ln\rho,\\
    R(\ue)&\equiv D(\ue)
       +\frac{1}{2}\ue\cdot\grad\ln\rho.
    \label{eq:DQR}
\end{align}
Then
\begin{equation}
    \divg\xin=\rho^{-1/2}Q(\ue),
    \qquad
    \delta\rho=-\rho^{1/2}R(\ue),
\end{equation}
and
\begin{equation}
    \delta P
    =-\rho^{1/2}
    \left[c_s^2 Q(\ue)+\ue\cdot\geff\right]
    \equiv -\rho^{1/2} S(\boldsymbol{u}).
    \label{eq:deltaP_weighted}
\end{equation}
This transformation is useful because the non-rotating restoring operator can be written in a symmetric weak form.

The momentum equation~\ref{eq:momentum_fullgravity}, divided by $\rho^{1/2}$, can 
then be written in the form 
\begin{align}
&
\underbrace{
-\nabla S(\boldsymbol{u})
-\frac{1}{2}S(\boldsymbol{u})\nabla\ln\rho
+R(\boldsymbol{u})\boldsymbol{g}_{\rm eff}
}_{\mathcal{L}_{\rm h}(\boldsymbol{u})}
\nonumber\\
&\hspace{2cm} \qquad
-\sigma^2\boldsymbol{u}
+\sigma
\underbrace{
\left(
-2{\rm i}\,\boldsymbol{\Omega}\times\boldsymbol{u}
\right)
}_{\mathcal{L}_{\rm C}(\boldsymbol{u})}
+\rho^{1/2}\nabla\phi
=0,
\label{eq:strong_form}
\end{align}
where $\phi\equiv\delta\Phi$, 
$\mathcal{L}_{\rm h}$ and $\mathcal{L}_{\rm C}$ are the hydrodynamic restoring and Coriolis operators, respectively. When we explore the consequences of adopting the Cowling approximation \citep{Cowling1941}, we drop the last term in Eq.~\ref{eq:strong_form}.

\subsection{Real helical decomposition for a single azimuthal wavenumber}

We write the cylindrical polar components of the displacement as $\boldsymbol{u}=(u_s,u_\phi,u_z)$. To keep the discrete matrices in the resulting FEM formulation real, we introduce helical variables for ($u_s,u_\phi$) while we keep $u_z$ unchanged, such that
\begin{equation}
    u_\phi=\ii w,
    \qquad
    u_+=u_s-w,
    \qquad
    u_-=u_s+w,
    \label{eq:helical}
\end{equation}
which involve only two independent quantities, 
\begin{equation}
    u_s=\frac{u_++u_-}{2},
    \qquad \textrm{and} \qquad
    w=\frac{u_--u_+}{2}.
    \label{eq:helical_inverse}
\end{equation}
The divergence operator, $\nabla\cdot\boldsymbol{u}$, for a fixed $m$, is expressed as:
\begin{equation}
\begin{split}
    D_m(\ue)=&\,
    \frac{1}{2}\left(\partial_s u_++\partial_s u_-\right)
    +\frac{(m+1)u_++(1-m)u_-}{2s}
    +\partial_z u_z.
\end{split}
    \label{eq:Dm}
\end{equation}
The apparent singularity suggested by the presence of $1/s$ terms here does not cause problems, because these terms are regular provided that the helical components have the correct behaviour on the axis $s=0$,
\begin{equation}
    u_+\propto s^{|m+1|},
    \qquad
    u_-\propto s^{|m-1|},
    \qquad
    u_z\propto s^{|m|},
    \quad \textrm{as} \quad s\rightarrow0.
    \label{eq:axis_regularity}
\end{equation}
These conditions determine which components vanish on the rotation axis and are imposed directly in the finite-element space.
 
The simple axis regularity conditions in Eq.~\eqref{eq:axis_regularity} make the helical variables particularly convenient for a real-valued finite-element discretization in the meridional $(s,z)$ domain.

\subsection{Boundary conditions on the gravitational potential}

The gravitational potential perturbation and its first derivative is represented continuously across the common star--vacuum interface. The normal-flux condition associated with the perturbed finite-density interface follows naturally from the weak Poisson equation and we do not need to impose it as a separate boundary condition.

At the circular (meridional slice of a spherical) exterior boundary, we impose
\begin{equation}
 \delta\Phi(R_{\Phi})=0,
 \qquad R_{\Phi}=4R_{\rm eq},
\end{equation}
which approximates a spatial decay to infinity if $R_\phi$ is sufficiently large. For $|m|>0$, scalar regularity additionally requires $\delta\Phi=0$ on the rotation axis; for $m=0$, the regular solution satisfies the corresponding natural axis condition. 

\subsection{From the strong form to the weak form in FEM}


To obtain the finite-element formulation, we recast the governing equations in weak form. We start from the strong-form momentum equation, Eq.~\eqref{eq:strong_form}, which holds in the stellar domain
$\mathcal{D}_\star$, and the perturbed Poisson equation,
Eq.~\eqref{eq:poisson_distribution}, which holds on the combined star--vacuum domain
$\mathcal{D}_\star\cup\mathcal{D}_{\rm vac}$.
The equations are multiplied by the corresponding test functions, $\boldsymbol{v}^{*}$ for the displacement and $\psi^{*}$ for the gravitational potential, and integrated over their respective domains. Integration by parts is applied where required to reduce the order of spatial derivatives. The correspondence between the strong-form terms
and the resulting weak bilinear forms is summarized in
Table~\ref{tab:strong_to_weak}.

\begin{table}
\centering
\caption{Correspondence between terms in the strong equations and the
bilinear forms used in the weak formulation. Here, IBP denotes integration
by parts.}
\label{tab:strong_to_weak}
\small
\renewcommand{\arraystretch}{1.25}
\begin{tabular}{@{}lll@{}}
\hline
Strong-form term & Test/domain & Weak form \\
\hline
$\mathcal{L}_{\rm h}(\boldsymbol{u})$
&
$\boldsymbol{v}$, $\mathcal{D}_\star$
&
$a_{\rm h}(\boldsymbol{v},\boldsymbol{u})$
\\

$\sigma\mathcal{L}_{\rm C}(\boldsymbol{u})$
&
$\boldsymbol{v}$, $\mathcal{D}_\star$
&
$\sigma a_1(\boldsymbol{v},\boldsymbol{u})$
\\

$-\sigma^2\boldsymbol{u}$
&
$\boldsymbol{v}$, $\mathcal{D}_\star$
&
$-\sigma^2 b(\boldsymbol{v},\boldsymbol{u})$
\\

$\rho^{1/2}\nabla\phi$
&
$\boldsymbol{v}$, $\mathcal{D}_\star$
&
$c(\boldsymbol{v},\phi)$
\\

$\nabla\!\cdot(\rho^{1/2}\boldsymbol{u})$
&
$\psi$, $\mathcal{D}_\star\cup\mathcal{D}_{\rm vac}$; IBP
&
$c(\boldsymbol{u},\psi)$
\\

$\nabla^2\phi$
&
$\psi$, $\mathcal{D}_\star\cup\mathcal{D}_{\rm vac}$; IBP
&
$k_\Phi(\psi,\phi)$
\\
\hline
\end{tabular}
\end{table}

In the Poisson equation, $\rho^{1/2}\boldsymbol{u}$ is extended by zero into the vacuum, so that its contribution is non-zero only in the stellar interior, while its distributional divergence also accounts for the finite-density stellar interface. A detailed derivation of the individual
bilinear forms is given in Appendix B.

The weak version of the equation governing the displacement is
\begin{equation}
a_{\rm h}(\boldsymbol{v},\boldsymbol{u})
+
\sigma a_1(\boldsymbol{v},\boldsymbol{u})
-
\sigma^2 b(\boldsymbol{v},\boldsymbol{u})
+
c(\boldsymbol{v},\phi)
=0,
\end{equation}
and the weak version of the Poisson equation is
\begin{equation}
c(\boldsymbol{u},\psi)
+
k_\Phi(\psi,\phi)
=0.
\end{equation}
These can be combined to obtain the mixed weak problem
\begin{align}
&
a_{\rm h}(\boldsymbol{v},\boldsymbol{u})
+
c(\boldsymbol{v},\phi)
+
c(\boldsymbol{u},\psi)
+
k_\Phi(\psi,\phi)
\nonumber\\
&\hspace{3cm} \qquad
+
\sigma a_1(\boldsymbol{v},\boldsymbol{u})
-
\sigma^2 b(\boldsymbol{v},\boldsymbol{u})
=0,
\label{mixed}
\end{align}
for all admissible test functions $(\boldsymbol{v},\psi)$. 

We can apply our finite-element discretization to Eq.~\ref{mixed}. The bilinear forms then correspond to the matrix blocks
\begin{equation}
a_{\rm h}\rightarrow A_{\rm h},
\qquad
a_1\rightarrow A_1,
\qquad
b\rightarrow B,
\end{equation}
and
\begin{equation}
c(\boldsymbol{v},\phi)\rightarrow C^T,
\qquad
c(\boldsymbol{u},\psi)\rightarrow C,
\qquad
k_\Phi\rightarrow K_\Phi.
\end{equation}
The discrete mixed eigenvalue problem is therefore
\begin{equation}
\begin{pmatrix}
A_{\rm h}+\sigma A_1-\sigma^2B & C^T\\
C & K_\Phi
\end{pmatrix}
\begin{pmatrix}
U\\
\Phi
\end{pmatrix}
=0.
\end{equation}
Equivalently, defining $X=(U,\Phi)^T$, 
and
\begin{equation}
\mathcal{A}_0=
\begin{pmatrix}
A_{\rm h} & C^T\\
C & K_\Phi
\end{pmatrix},
\quad
\mathcal{A}_1=
\begin{pmatrix}
A_1 & 0\\
0 & 0
\end{pmatrix},
\quad
\mathcal{A}_2=
\begin{pmatrix}
-B & 0\\
0 & 0
\end{pmatrix},
\end{equation}
the problem can be represented in the quadratic polynomial form
\begin{equation}
\left(
\mathcal{A}_0
+\sigma\mathcal{A}_1
+\sigma^2\mathcal{A}_2
\right)X=0.
\end{equation}
This is the problem that we will solve in this work.

The two-dimensional oscillation domain is discretized in the meridional $(s,z)$ plane using an unstructured triangular mesh generated with Gmsh \citep{geuzaine2009gmsh} and
subsequently imported into DOLFINx \citep{baratta2025}.  The stellar boundary is defined by the
finite-enthalpy isosurface corresponding to the same fractional enthalpy,
$r_{\rm cut}/R=0.99$, as in the non-rotating $N=3$ Lane--Emden model.
A target Gmsh element size of $h=0.01$ was
adopted, corresponding to a characteristic spatial resolution of approximately $1\%$ of $R_{\rm eq}$.  The resulting half-meridional domain contains 35\,759 triangular cells.  The initial Gmsh discretization contains
18\,254 mesh vertices and 36\,389 elements in total, including approximately 630 one-dimensional boundary elements on the stellar surface and rotation axis.  

The quadratic polynomial eigenvalue problem is solved using the SLEPc \citep{slepc2005,Petsc2026} 
Polynomial Eigenvalue Problem (PEP) interface, with the assembled operators
stored in PETSc sparse-matrix format. Computing the full spectrum is very
expensive in terms of both memory usage and computational cost. We therefore
use targeted searches, requesting 30 eigenpairs near each prescribed
rotating-frame target frequency, $\sigma_{\rm targ}$.


An important feature of the 
formulation is that $\mathcal{A}_2$ is singular, so the polynomial eigenvalue problem contains formally infinite eigenvalues arising from the gravitational potential degrees of freedom, which do not have an associated inertial term. We use targeted SLEPc searches to compute the finite eigenvalues near $\sigma_{\rm targ}$; the corresponding eigenvectors contain degrees of freedom for both the displacement and the gravitational potential.

For each computed eigenpair $(\sigma,X)$, we evaluate the normalized polynomial residual
\begin{equation}
\epsilon_{\rm EVP}
=
\frac{
\left\|
\left(
\mathcal A_0
+\sigma\mathcal A_1
+\sigma^2\mathcal A_2
\right)X
\right\|
}{
\left(
\|\mathcal A_0\|
+|\sigma|\,\|\mathcal A_1\|
+|\sigma|^2\|\mathcal A_2\|
\right)\|X\|
},
\end{equation}
which provides a direct numerical measure of the accuracy of the computed eigenpair. The residual is typically $\epsilon_{\rm EVP}\sim10^{-13}$ or smaller for the modes we present in this work, indicating that the discretized problem has been solved accurately.  
\section{Validation against non-rotating spherical predictions} 

The non-rotating limit $q=0$ provides an independent validation of our model. For a spherical equilibrium stellar model, we consider linearised adiabatic non-radial oscillations, and use the standard Dziembowski-style variables \citep{Dziembowski1971},
\begin{equation}
 y_1=\frac{\xi_r}{r},
 \qquad
 y_2=\frac{\delta P/\rho+\delta\Phi}{gr},
 \qquad
 y_3=\frac{\delta\Phi}{gr},
 \qquad
 y_4=\frac{1}{g}\frac{d\delta\Phi}{dr},
\end{equation}
where $g=GM_r/r^2$ is the magnitude of the gravitational acceleration. The adiabatic equations can then be written \citep[e.g.][]{Unno1989,Townsend2013}
\begin{align}
 r\frac{dy_1}{dr}
 &=
 (V_g-3)y_1
 +\left[
 \frac{\ell(\ell+1)}{c_1\omega^2}-V_g
 \right]y_2
 +V_g y_3 ,
 \\
 r\frac{dy_2}{dr}
 &=
 (c_1\omega^2-A^\ast)y_1
 +(A^\ast-U+1)y_2
 -A^\ast y_3 ,
 \\
 r\frac{dy_3}{dr}
 &=
 (1-U)y_3+y_4 ,
 \\
 r\frac{dy_4}{dr}
 &=
 UA^\ast y_1
 +UV_g y_2
 +\left[\ell(\ell+1)-UV_g\right]y_3
 -Uy_4 ,
\end{align}
with
\begin{equation}
 c_1=\frac{r^3}{M_r},
 \qquad
 V_g=\frac{gr}{c_s^2},
 \qquad
 U=\frac{d\ln M_r}{d\ln r},
 \qquad
 A^\ast=\frac{rN^2}{g},
\end{equation}
when using units such that $G=M=R_{\rm eq}=1$.

At the finite stellar cutoff radius, the free surface condition is $\delta P + \boldsymbol{\xi}\cdot \nabla P=0$, which corresponds to the Lagrangian pressure condition $y_2-y_3-y_1=0$. The gravitational potential and its derivative are then matched to a vacuum solution satisfying Laplace's equation between
$r=r_{\rm cut}$ and $R_{\Phi}=4$, with
$\delta\Phi(R_{\Phi})=0$. Thus, the one- and two-dimensional
calculations use the same fluid cutoff and the same exterior
gravitational boundary, which allows us to make a direct numerical comparison.


Agreement with the independent one-dimensional calculation provides an
important validation of the mixed displacement--potential FEM formulation.
Using the same fluid cutoff radius, $r_{\rm cut}$, and exterior potential
domain, $R_{\Phi}$, the one-dimensional calculation gives
$\sigma_{\ell=3,g_1}=2.60134109$ and
$\sigma_{\ell=2,p_1}=3.90698854$ (both with radial order $n=1$). For the corresponding $q=0$ model
constructed with the full two-dimensional SCF procedure and subsequently
solved with the FEM code, we obtain $2.60141113$ and $3.90748682$,
respectively. These correspond to relative differences of approximately
$2.7\times10^{-5}$ and $1.3\times10^{-4}$.

As an additional test, we map the spherical $N=3$ model directly onto the two-dimensional FEM mesh, bypassing the SCF reconstruction. The resulting low-degree mode frequencies agree with the one-dimensional calculation at the sub-ppm level. We retain the $q=0$ SCF model as the main end-to-end test, since it follows the same procedure used for the rotating models; its slightly larger frequency differences mainly reflect errors in constructing and mapping the equilibrium model.

The assembled restoring matrix is symmetric to numerical accuracy, and well-resolved eigenpairs typically have normalized residuals of $\sim10^{-15}$--$10^{-13}$, indicating that they each provide accurate solutions of the discrete system. Together with the frequency comparison, these tests independently probe the hydrodynamic restoring operator, the symmetric displacement--potential coupling, and the vacuum Poisson solve.

\section{Comparison with the 2D mode calculations of Reese} 
\label{subsec:reese_fullgravity}

\begin{figure*}
\centering
\includegraphics[width=1.0\textwidth]{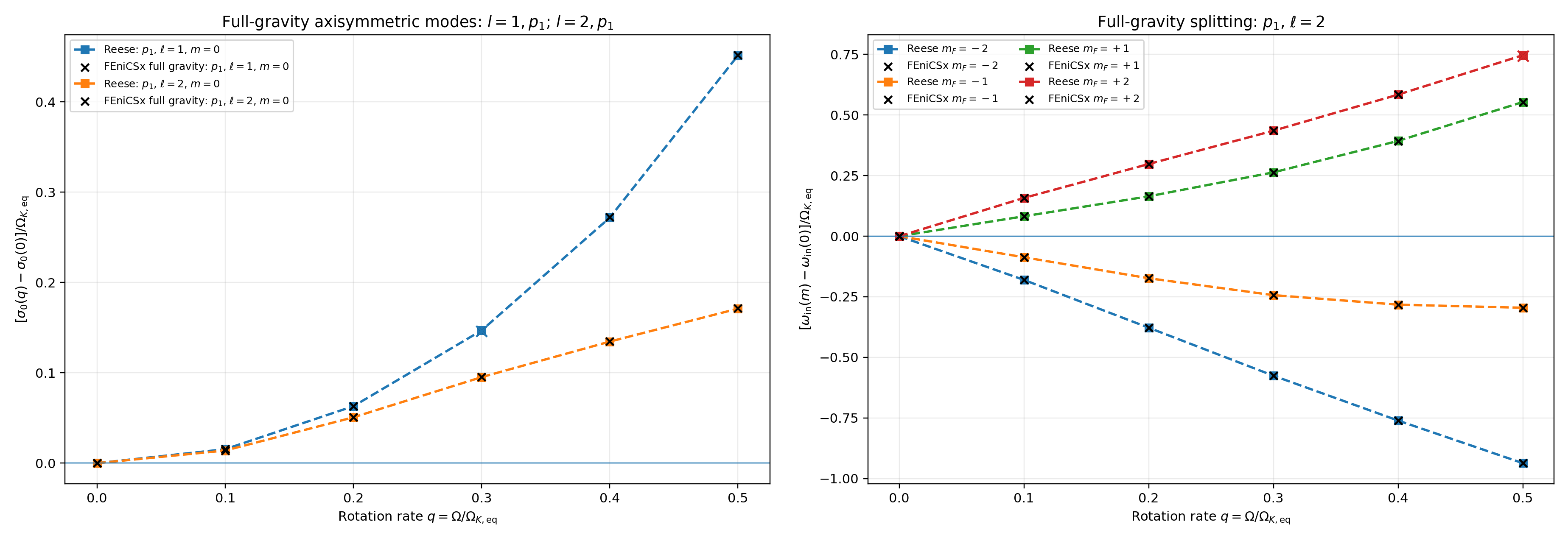} 
\caption{ 
Comparison of the present FEniCSx calculations with the two-dimensional TOP code frequencies of \citet{Reese2006} for the
$p_1$ modes of a uniformly rotating $n=3$ polytrope.
\textit{Left:} rotation-induced frequency shifts,
$[\sigma_0(q)-\sigma_0(0)]/\Omega_{K,\rm eq}$, for the axisymmetric
$\ell=1$ and $\ell=2$ branches.
\textit{Right:} inertial-frame rotational splittings,
$[\omega_{\rm in}(m)-\omega_{\rm in}(0)]/\Omega_{K,\rm eq}$, for the
$\ell=2$ multiplet with $m=-2,-1,+1,+2$, where
$\omega_{\rm in}=\sigma+m\Omega$.
Dashed curves with square symbols show the Reese/TOP results, while crosses show the present FEniCSx frequencies. The close agreement in both panels demonstrates that the FEM calculation reproduces the axisymmetric rotational shifts as well as the non-axisymmetric multiplet splitting over $0\le q\le0.5$.}
\label{fig:fig5_p1}
\end{figure*}

\citet{Reese2006a} calculated the free modes of a uniformly rotating $N=3$ polytrope using a two-dimensional multi-domain spectral method (implemented in the TOP code). This accounts for non-perturbative effects of rotation, including the rotational deformation and the associated perturbations to the gravitational potential. He tabulated rotating-frame 
eigenfrequencies for uniformly rotating $N=3$ polytropes in his Appendix~F. The tabulated rotation and pulsation
 frequencies are normalized using the polar
radius $R_{\rm pol}$, whereas our SCF models are normalized to $R_{\rm eq}$. The conversion (from $q_\mathrm{pol}$, using the polar radius) to the present normalization (without subscripts on $q$) is 
\begin{equation}
    q
    =
    \frac{q_{\rm pol}}
    {(1-\epsilon)^{3/2}},
    \label{eq:reese_q_conversion}
\end{equation}
where $\epsilon$ is the flattening parameter,
$\epsilon= 1-\frac{R_{\rm pol}}{R_{\rm eq}}$.
Similarly, for the rotating-frame pulsation frequency,
\begin{equation}
    \sigma_{\rm eq}
    =
    \frac{\sigma_{\rm pol}}
    {(1-\epsilon)^{3/2}},
    \label{eq:reese_sigma_conversion}
\end{equation}
where $\sigma_\mathrm{eq}$ and $\sigma_{\rm pol}$ are the rotating-frame oscillation frequencies normalized by $R_{\rm eq}$ and $R_{\rm pol}$, respectively. The frequencies from Reese were first converted using
Eq.~(\ref{eq:reese_sigma_conversion}) and then interpolated as a
function of the equatorial rotation parameter $q$ to the sequence
$q=0,0.1,\ldots,0.5$\footnote{The tabulated frequencies of \citet{Reese2006a} extend only to $\Omega/\sqrt{GM/R_{\rm eq}^3}=0.58$. Thus, we do not have reference values from their calculations for higher rotation rates and cannot
make a direct comparison for $q>0.58$.} used in the present calculations. The two calculations also adopt opposite conventions for the azimuthal order, so we use the conversion: $ m_{\rm Reese}=-m_{\rm FEM}$.

When comparing the rotating-frame frequencies from FEniCSx with the converted Reese/TOP frequencies, both sets of values are in close agreement with each another over the full range of $q$ we consider. The maximum absolute fractional difference for the $p_1$, $\ell=2$ modes is approximately $3.0\times10^{-4}$.
This provides a comprehensive test of our FEM implementation, as well as an independent verification of the results obtained by \citet{Reese2006a} and \citet{Ouazzani2012}.



We find that the frequency differences between our FEM and Reese's increase systematically with rotation for higher $\ell$ modes. For example, the $p_2$, $\ell=3$ multiplet exhibits a maximum rotation-dependent difference of approximately $1.0\times10^{-3}$ at $q=0.5$. This behaviour is consistent with the increasing sensitivity of higher-radial-order acoustic modes to the outer stellar layers. In particular, the present calculation imposes $\Delta P=0$ on a finite-enthalpy surface corresponding approximately
to $r/R=0.99$ in the non-rotating model, whereas the Reese/TOP calculation treats the true polytropic surface 
The transfer of the SCF background onto the independent finite-element mesh also contributes an additional numerical error, which becomes increasingly important for shorter-wavelength (higher $\ell$) eigenfunctions. 

To separate the differences induced by rotation 
from the small, $10^{-4}$-level discretisation and modelling errors, 
we also define
\begin{equation}
    \Delta_{\rm rot}(q)
    =
    \frac{
    \sigma_{\rm FEM}(q)/\sigma_{\rm FEM}(0)
    }{
    \sigma_{\rm Reese}(q)/\sigma_{\rm Reese}(0)
    }
    -1 ,
    \label{eq:reese_rot_residual}
\end{equation}
where subscripts FEM and Reese denote the two different predictions, respectively. By construction, $\Delta_{\rm rot}(0)=0$. For the $\ell=2,p_1$, $\ell=3,p_1$, $\ell=1,p_2$, and $\ell=2,p_2$ mode families, the maximum values of $|\Delta_{\rm rot}|$ are $\sim (1$--$4)\times10^{-4}$.
 Thus, for the clean low-order modes, the rotational
frequency evolution obtained with the two independent
two-dimensional methods agrees at the level of a few parts in
$10^{4}$.

Figure~\ref{fig:fig5_p1} provides a compact comparison between
the present 
FEniCSx FEM calculation and the results of
Reese~\citep{Reese2006a} for the $p_1$ family \citep[in a similar way to Fig.~5 of][who used the spectral ACOR code]{Ouazzani2012}. The left-hand panel shows the rotation-induced frequency shifts of the axisymmetric ($m=0$) $\ell=1$ and $\ell=2$ modes, $(\sigma_0(q)-\sigma_0(0))/\Omega_{K,\rm eq}$,
as a function of
$q=\Omega/\Omega_{K,\rm eq}$, where $\Omega_{K,\rm eq}$ is the critical (Keplerian) rotational velocity at the equator, which is the same as the dynamical frequency $\Omega_{\rm dyn}$. For both branches, the frequencies from FEniCSx are in excellent agreement with the Reese/TOP data over the full range $0\le q\le0.5$, including the pronounced nonlinear increase in the frequency of the $\ell=1$ branch at larger rotation rates. The right-hand panel provides a more stringent non-axisymmetric test using the complete $\ell=2$, $p_1$ multiplet. Here we plot the inertial-frame splitting $(\omega_{\rm in}(m,q)-\omega_{\rm in}(0,q))/\Omega_{K,\rm eq}$, where $\omega_{\rm in}$ is $\sigma+m\Omega$, 
for $m=-2,-1,+1,+2$ using our sign conventions as implemented in FEniCSx.
The points from the FEniCSx predictions lie very close to the
corresponding Reese branches (with any differences being much smaller than the sizes of the symbols in the figure), and reproduce both the sign and magnitude of the prograde--retrograde asymmetry and its nonlinear dependence on rotation. The simultaneous agreement of the axisymmetric centroid shifts and the non-axisymmetric multiplet structure provides a sensitive validation of our implementation of the centrifugal deformation, Coriolis coupling, and perturbed self-gravity in this two-dimensional FEM formulation.

Overall, the comparison with \citet{Reese2006a} demonstrates that the finite-element formulation reproduces both the centroid shifts and the
non-axisymmetric multiplet structure of rapidly rotating
self-gravitating polytropes. For the low-order, low-$\ell$ mode ($\ell \le 3$) families, agreement is obtained at the $O(10^{-4})$ 
level in the rotation-dependent frequencies for a substantial fraction of the rapid-rotation regime. The larger differences found for higher-order
or strongly mixed modes are correlated with reduced parent-harmonic
purity, avoided crossings, and increased sensitivity to the location and accurate modelling of the finite outer fluid boundary. These results provide an external validation of the complete two-dimensional displacement--potential finite-element eigenproblem that we have presented.

\section{Mode identification and tracking with rotation}
\label{sec:eigenfunctions}

\subsection{Eigenfunction reconstruction}

It is instructive to analyse the eigenfunctions of rapidly rotating models to determine how they differ from non-rotating or weakly rotating ones. Each finite-element eigenvector is stored using real helical variables $(u_+,u_-,u_z)$, from which the cylindrical components are reconstructed using Eq.~\eqref{eq:helical_inverse}. The corresponding spherical polar components in a meridional plane are
\begin{align}
    u_r
    &=\frac{s u_s+z u_z}{r},\\
    u_\theta
    &=\frac{z u_s-s u_z}{r},
    \label{eq:spherical_components}
\end{align}
where $r=(s^2+z^2)^{1/2}$. The real physical displacement is
\begin{equation}
\begin{split}
    \xin_{\rm phys}(s,z,\phi,t)
    =\rho^{-1/2}\,\Re\Big[
    \left(
      u_s\ve{e}_s+u_z\ve{e}_z+\ii w\ve{e}_\phi
    \right)
    e^{\ii m\phi-\ii\sig t}
    \Big].
    \label{eq:physical_displacement}
\end{split}
\end{equation}
Thus, the meridional components are $90^\circ$ out-of-phase with the stored azimuthal component, $w$. 


To characterize the angular structure, we project the numerically computed
eigenfunction onto vector spherical harmonics on a set of surface-fitted
interior shells labelled by the coordinate $\zeta$ in Eq.~\eqref{r_zeta},
\begin{equation}
\begin{aligned}
    \ue(\zeta,\theta,\phi)
    = \sum_{\ell\ge |m|}
    \Big[
        &u_{r,\ell}(\zeta)Y_\ell^m\,\ve{e}_r
        +u_{h,\ell}(\zeta)\grad_hY_\ell^m
        \\
        &\qquad \qquad \qquad +u_{t,\ell}(\zeta)
        \ve{e}_r\times\grad_hY_\ell^m
    \Big],
\end{aligned}
\label{eq:vsh_decomposition}
\end{equation}
where the first two terms represent the radial and horizontal
spheroidal components, respectively, while the final term is the
toroidal component. Here, $\nabla_h \equiv \boldsymbol{e}_\theta\,\partial_\theta
+\boldsymbol{e}_\phi\,(\sin\theta)^{-1}\partial_\phi$ denotes the dimensionless horizontal gradient on the unit sphere. We define the harmonic ``purity'' as the fraction of the projected squared-amplitude norm associated with the dominant spherical harmonic degree $\ell_{\rm dom}$. This is a useful and reliable diagnostic for isolated modes at slow and moderate rotation rates, but it should not be interpreted as an exact quantum number once the centrifugal deformation and Coriolis acceleration strongly couple different harmonic degrees.

The equatorial parity provides an additional exact classification for
uniform rotation. Since
\begin{equation}
    Y_\ell^m(\pi-\theta,\phi)
    =
    (-1)^{\ell+|m|}
    Y_\ell^m(\theta,\phi),
\end{equation}
the radial and meridional components of a spheroidal vector spherical
harmonic satisfy
\begin{align}
    u_r^{\rm sph}(\pi-\theta)
    &=
    (-1)^{\ell+|m|}
    u_r^{\rm sph}(\theta),
    \nonumber\\
    u_\theta^{\rm sph}(\pi-\theta)
    &=
    -(-1)^{\ell+|m|}
    u_\theta^{\rm sph}(\theta),
    \label{eq:parity_sph}
\end{align}
whereas for the toroidal vector spherical harmonic,
$\boldsymbol{e}_r\times\nabla_hY_\ell^m$,
\begin{align}
    u_\theta^{\rm tor}(\pi-\theta)
    &=
    (-1)^{\ell+|m|}
    u_\theta^{\rm tor}(\theta),
    \nonumber\\
    u_\phi^{\rm tor}(\pi-\theta)
    &=
    -(-1)^{\ell+|m|}
    u_\phi^{\rm tor}(\theta).
    \label{eq:parity_tor}
\end{align}
Thus spheroidal and toroidal contributions belonging to the same
overall equatorial-parity family occupy opposite $\ell$-parity sectors.
Modes belonging to opposite parity families cannot mix in a uniformly
rotating equilibrium with reflection symmetry about the equator. This
provides a useful constraint for mode tracking in dense spectra, since
candidate matches of the opposite parity can be excluded even when their
frequencies are very close.



\begin{figure*}
\centering
\includegraphics[width=0.88\textwidth]{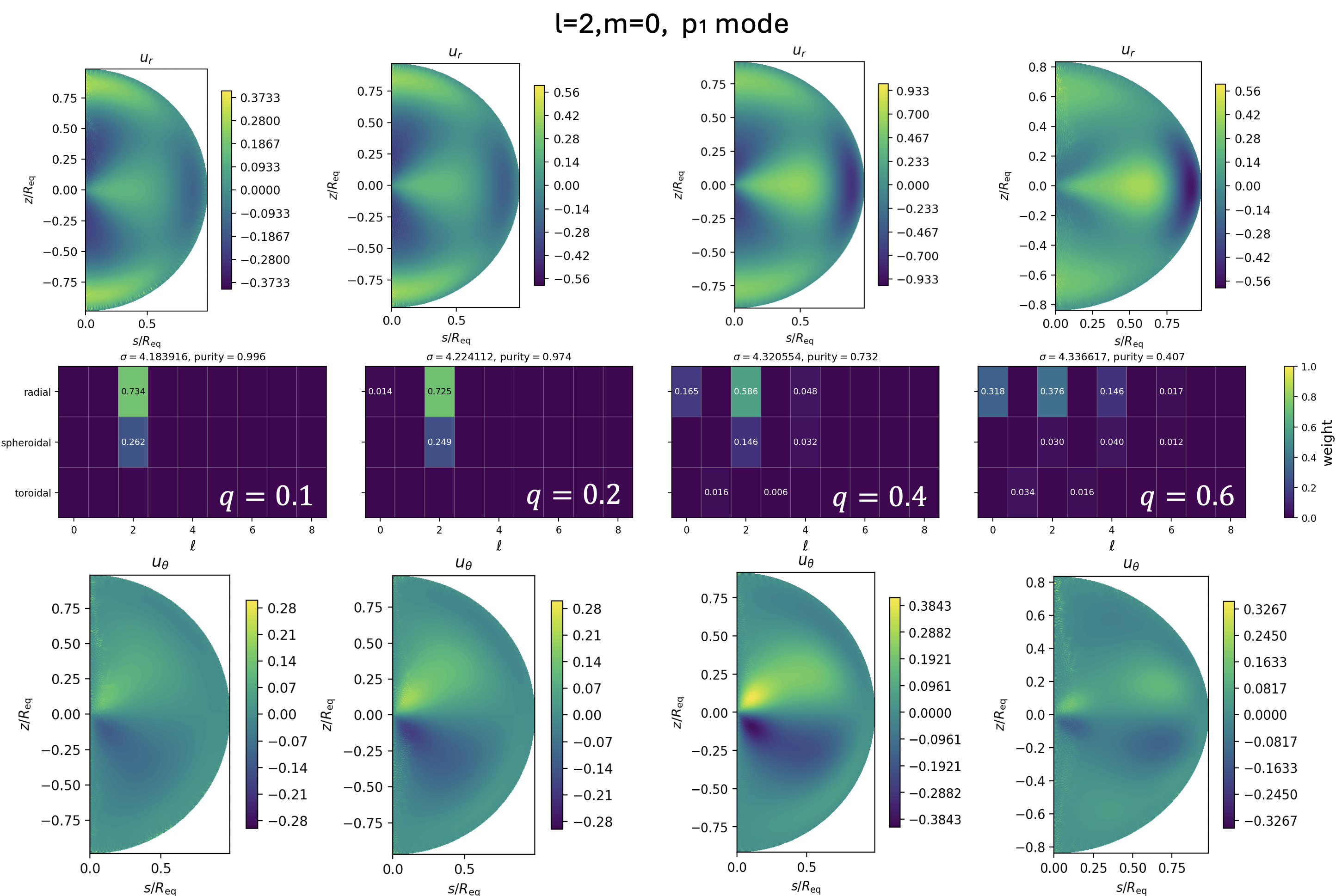}
\caption{Evolution of the density-weighted displacement for the tracked $\ell=2$, $m=0$, $p_1$ branch at $q=0.1$, $0.2$, $0.4$, and $0.6$. The upper row shows $u_r$ and the lower row $u_\theta$. The exact equatorial parity is preserved throughout: $u_r$ is even and $u_\theta$ is odd about the equator. The middle panel illustrates the spectral components of the mode, in the form of $( E_{\ell,\rm radial},E_{\ell,\rm spheroidal},E_{\ell,\rm toroidal})$. As rotation increases, the stellar boundary becomes increasingly oblate and the mode departs from a single $Y_2^0$ geometry through coupling to spherical harmonics with the same parity (e.g., $\ell=0,2,4, ...$ and $\ell=1,3,5...$, seen in the middle panel). Since eigenvectors have arbitrary normalization, the panels should be compared through their spatial morphology rather than their absolute plotted amplitudes.}
\label{fig:eigenfun1}
\end{figure*}

\begin{figure*}
\centering
\includegraphics[width=0.88\textwidth]{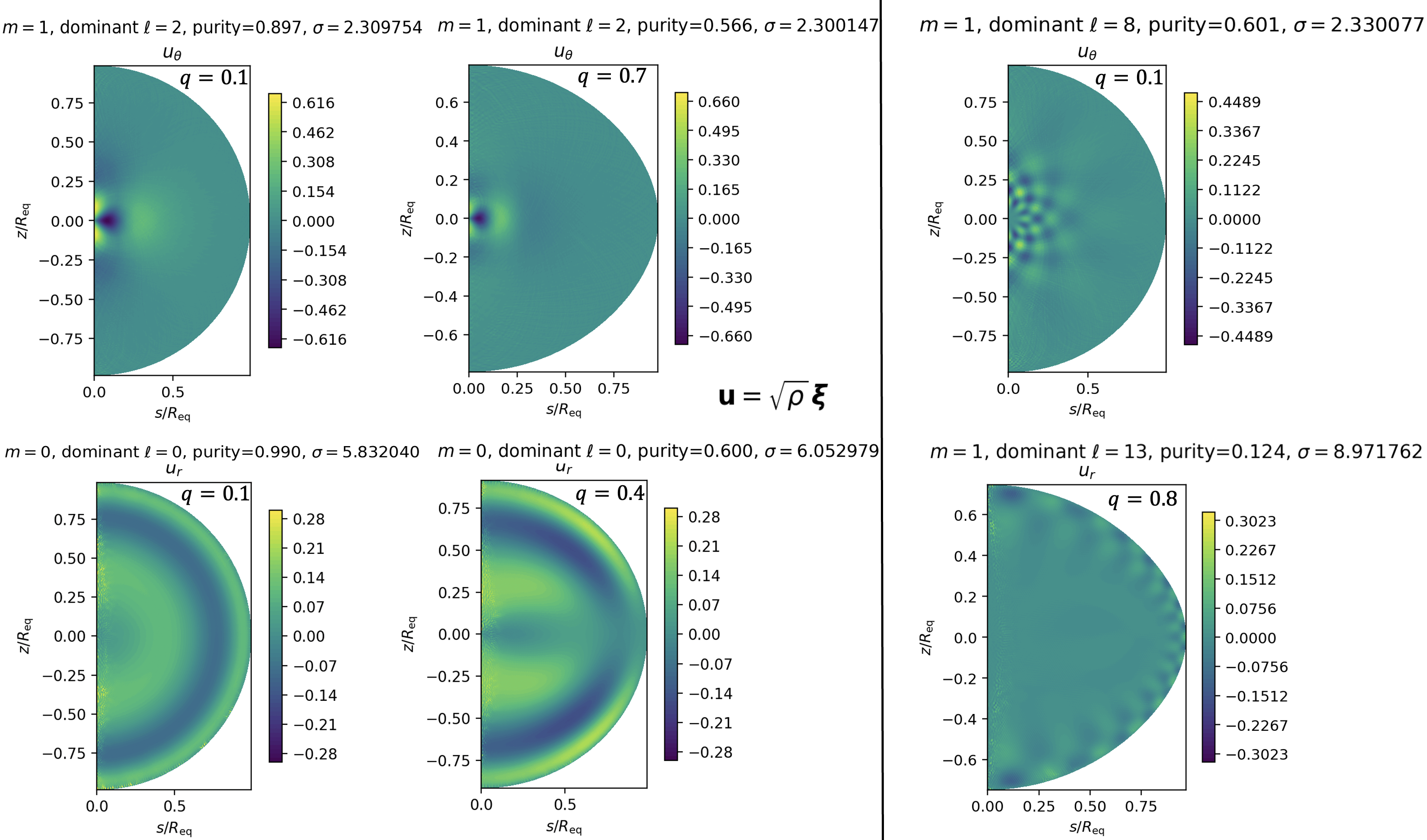}
\caption{
Examples of oscillation modes computed with our 2D FEniCSx implementation. 
All panels show meridional cuts of selected components of 
$\mathbf{u}=\sqrt{\rho}\,\boldsymbol{\xi}$, with rotation rate 
$q=\Omega/\Omega_{\mathrm{K,eq}}$. 
The upper-left and upper-middle panels show the same low-order 
$\ell_{\mathrm{dom}}=2$, $m=1$ g-mode branch at $q=0.1$ and $0.7$, 
respectively. 
The lower-left and lower-middle panels show an $m=0$ radial acoustic 
mode at $q=0.1$ and $0.4$, illustrating the increasing rotational coupling to higher even spherical harmonic degrees. 
The upper-right panel shows a higher-degree g-mode candidate, while the lower-right panel shows a high-frequency whispering-gallery acoustic mode, 
with the oscillation concentrated near the stellar surface.}
\label{fig:modes_pg}
\end{figure*}


\subsection{Method for mode tracking with rotation}
\label{subsec:mode_tracking}

To follow individual oscillation modes as the rotation rate is increased, we
start from the non-rotating, $m$-degenerate eigenmodes and use frequency
continuation to identify their rotating counterparts. At each new value of
$q$, a predicted frequency, $\sigma_{\rm pred}$, is estimated from the
previously tracked points, and only eigenmodes within a finite frequency
window around this prediction are considered. For axisymmetric ($m=0$)
modes, the first-order rotational correction vanishes, so that at sufficiently slow rotation
\begin{equation}
    \sigma(q)
    =
    \sigma_0+a_2q^2+a_4q^4+\cdots .
\end{equation}
We therefore extrapolate these branches as functions of $q^2$. For
non-axisymmetric ($m\neq0$) modes, the Coriolis force introduces a
first-order frequency shift proportional to $q$, so the continuation is
instead performed directly as a function of $q$. For gravity modes that become strongly modified by rotation, low-order
perturbative frequency scalings may no longer provide a reliable predictor.
We therefore use a local linear extrapolation from the two most recently
tracked points.

The above approach becomes unreliable once centrifugal distortion and Coriolis coupling produce dense spectra, avoided crossings, and strong mixing between different spherical harmonic components. We therefore
characterize each mode by an angular ``fingerprint'' constructed from the fractional squared-amplitude contributions of its radial, spheroidal, and toroidal components. We retain spherical harmonic degrees $\ell=0,\ldots,16$, this gives a 51-component fingerprint,
\begin{equation}
    {\boldsymbol E}
    =
    \left(
    E_{r,0},\ldots,E_{r,16},
    E_{{\rm sph},0},\ldots,E_{{\rm sph},16},
    E_{{\rm tor},0},\ldots,E_{{\rm tor},16}
    \right),
\label{eq:E16}
\end{equation}
normalized such that $\sum_k E_k=1$. Here, each $E_k$ measures the relative angular power contained in the corresponding vector-spherical-harmonic
component. Figure~3 (middle panel) illustrates these fingerprints for an $\ell_{\rm dom}=2$, $m=0$ mode as the rotation rate increases from
$q=0.1$ to $q=0.6$.

To compare fingerprints between adjacent rotation rates, we define the
overlap
\begin{equation}
    {\cal B}_{ij}
    =
    \sum_k \sqrt{E_{i,k}E_{j,k}},
    \qquad
    0\le {\cal B}_{ij}\le 1,
    \label{Boverlap}
\end{equation}
where ${\cal B}_{ij}=1$ corresponds to identical normalized fingerprints. This measure is substantially more robust than following only the dominant spherical degree $\ell_{\rm dom}$, because a continuously evolving mode may exchange its dominant harmonic through an avoided crossing while retaining a similar overall distribution of angular and vector components.

The fingerprints $\boldsymbol E$ for each mode are calculated from a vector spherical harmonic decomposition of each two-dimensional FEM eigenfunction.  This decomposition is performed during the eigenmode classification stage. Because the equilibrium configuration is oblate, the
eigenfunction is sampled on a set of surface-fitted shells,
\begin{equation}
    r(\theta)=\zeta\,r_{\rm cut}(\theta),
    \label{r_zeta}
\end{equation}
with several fixed values of the normalized radial coordinate $\zeta$.
The components of the numerical eigenfunction $\boldsymbol{u}$ for each azimuthal wavenumber $m$ are projected onto the appropriate
radial, spheroidal, and toroidal vector spherical harmonic components (Eq. \ref{eq:vsh_decomposition}) on each shell. 
After division by the corresponding basis norms, the squared projection
amplitudes are normalized by the total angular energy on that shell,
\begin{equation}
    E_{\rm shell}
      =\int
      \left(u_r^2+u_\theta^2+w^2\right)\,{\rm d}\mu ,
\end{equation}
giving the fractional radial, spheroidal, and toroidal contributions
$E_{r,\ell}$, $E_{{\rm sph},\ell}$, and $E_{{\rm tor},\ell}$, respectively. 

Equatorial parity is an exact symmetry of a uniformly rotating,
axisymmetric equilibrium. For a mode originating from a non-rotating
spheroidal degree $\ell_0$, the radial and spheroidal components occupy
the same $\ell$-parity sector as $\ell_0$, whereas the
Coriolis-generated toroidal components occupy the opposite sector.
We therefore define the parity fraction
\begin{equation}
    P_{\rm parity}
    =
    \frac{
    \displaystyle
    \sum_{\ell\equiv\ell_0\,({\rm mod}\,2)}
    \left(E_{r,\ell}+E_{{\rm sph},\ell}\right)
    +
    \sum_{\ell\not\equiv\ell_0\,({\rm mod}\,2)}
    E_{{\rm tor},\ell}
    }{
    E_{\rm tot}
    },
\end{equation}
which measures the fraction of the projected squared-amplitude norm contained in the expected parity sectors. In the continuum limit, $P_{\rm parity}=1$, small departures from unity arise from numerical projection and the finite spherical harmonic truncation. In the mode tracking, we require $P_{\rm parity}\geq0.82$, and discard candidates
below this threshold.



As we advance from a current rotation rate $q_i$ to a new rotation rate $q_{i+1}$, we first predict the mode
frequency, $\sigma_{\rm pred}$, from the previously selected points.
Only eigenmodes satisfying
\begin{equation}
    |\sigma-\sigma_{\rm pred}| \leq \Delta\sigma
\end{equation}
are retained as candidates, with $\Delta\sigma=0.25$ in the
dimensionless dynamical frequency units used here. The remaining candidates are ranked using the score
\begin{equation}
    {\cal S}
    =
    w_\sigma
    \frac{|\sigma-\sigma_{\rm pred}|}{\Delta\sigma}
    +
    w_{\cal B}(1-{\cal B})
    +
    w_P(1-P_{\rm parity})
    +
    w_\ell(1-R_{\ell_0})
    \label{eq:tracking_score}
\end{equation}
where smaller values correspond to a closer continuation of the mode from $q_i$ to $q_{i+1}$. Here ${\cal B}$ is the overlap between the 51-component angular fingerprints at the two successive rotation rates in Eq.~\eqref{Boverlap}, and $P_{\rm parity}$ measures the fraction of the projected norm in
the expected equatorial parity sectors. The final term measures how strongly the candidate retains the
spherical harmonic degree $\ell_0$ of its non-rotating parent. We define
\begin{equation}
    P_{\ell_0}
    =
    E_{r,\ell_0}
    +
    E_{{\rm sph},\ell_0}
    +
    E_{{\rm tor},\ell_0},
\end{equation}
and
\begin{equation}
    R_{\ell_0}
    =
    \min\left[
    \frac{P_{\ell_0}(q_{i+1})}
         {P_{\ell_0}(q_i)},
    1
    \right].
\end{equation}
Thus, a decrease in the parent-$\ell$ contribution is penalized, whereas an increase is not. This choice reflects the role of \(P_{\ell_0}\) as a parent-mode retention diagnostic: loss of the original \(\ell_0\) contribution is evidence against a continuation.
We adopt the weights:
\begin{equation}
    w_\sigma=2.0,\qquad
    w_B=2.0,\qquad
    w_P=0.30,\qquad
    w_\ell=0.20.
\end{equation}
Thus, frequency continuity and continuity of the full angular
fingerprint provide the principal discriminants, while parity,
retention of the parent harmonic, and instantaneous harmonic purity provide weaker additional constraints.

Among the candidates that satisfy the parity criterion, the mode with the lowest score is selected as the continuation to $q_{i+1}$, while several of the next-best candidates are retained for diagnostic purposes.

Importantly, the parent degree need not remain the dominant spherical harmonic along a continuously evolving branch. For example, along the tracked $\ell_0=2$, $p_1$, $m=0$ branch, the parent-$\ell$ fraction
decreases substantially as $q$ increases, while the fingerprint overlap
between successive rotation steps remains close to unity over much of
the sequence. We therefore interpret a change in $\ell_{\rm dom}$ as
evidence of rotational mode mixing rather than, by itself, a failure of
the tracking procedure. If the fingerprint overlap also decreases
sharply, however, the continuation becomes less secure, as may occur
near an avoided crossing or in a region where two candidate branches
become difficult to distinguish.


\subsection{Rotational effects on mode morphologies}

Figure~\ref{fig:eigenfun1} illustrates the continuous deformation, frequency evolution, and angular-mode mixing of the tracked $\ell=2$, $m=0$ $p_1$ branch as the rotation rate increases. At $q=0.1$, the radial displacement retains the smooth global quadrupolar morphology expected for its non-rotating $\ell=2$ parent, while $u_\theta$ shows the corresponding antisymmetry across the equator. Consistently, its angular-energy fingerprint is still almost entirely confined to the $\ell=2$ radial and spheroidal components, with an $\ell=2$ purity of $0.996$. 

At $q=0.2$, the same overall mode topology remains recognizable, although centrifugal flattening and latitudinal redistribution become more visible; the fingerprint likewise shows the first detectable leakage into the $\ell=0$ radial component, while the $\ell=2$ purity remains high at $0.974$.

The rotational mixing becomes much stronger by $q=0.4$: $u_r$ develops a pronounced equatorial interior lobe together with enhanced high-latitude structure, and $u_\theta$ becomes more strongly concentrated toward the equatorial region. Correspondingly, radial
energy is redistributed from $\ell=2$ into the even-$\ell$ components
$\ell=0$ and $4$, while the spheroidal part also develops a clear
$\ell=4$ contribution. At the same time, Coriolis coupling generates a
toroidal component with the complementary odd-$\ell$ sequence, principally
$\ell=1$ with a weaker $\ell=3$ contribution. 

By $q=0.6$, the stellar cavity is substantially oblate and the eigenfunction is no longer well represented by a single spherical harmonic. The radial fingerprint is then shared mainly among $\ell=0$, $2$, and $4$, with a weaker extension to $\ell=6$, whereas the spheroidal contribution is distributed over even $\ell=2,4,6$ and the
toroidal contribution over odd $\ell=1,3,5,\ldots$. This even-$\ell$
radial/spheroidal and odd-$\ell$ toroidal hierarchy is consistent with the
equatorial symmetry of the axisymmetric parent mode and the angular coupling introduced by rotation. 

Quantitatively, the projected $\ell=2$ fraction
decreases from $0.996$ at $q=0.1$ to $0.974$, $0.732$, and $0.407$ at
$q=0.2$, $0.4$, and $0.6$, respectively, while the tracked rotating-frame
eigenfrequency changes from
$\sigma_{\rm rot}=4.183916$ to $4.224112$, $4.320554$, and $4.336617$ over
the same sequence. Thus, the frequency initially increases appreciably with
rotation but its variation becomes much weaker at the largest rotation
rates, at the same time that the angular fingerprint broadens strongly.
Nevertheless, the fingerprint similarity to the immediately preceding model
remains approximately $0.99$, $0.98$, and $0.98$ across the latter steps.
The combined eigenfunction morphology and angular-energy decomposition
therefore demonstrate continuous centrifugal deformation and Coriolis-driven mode mixing along a single physical branch. This demonstrates the ability of our method to track the evolution of a particular mode as a parameter (in this case, the rotation rate $q$) is varied by introducing a mode ``fingerprint'' that is used to calculate a ``score'' (including information such as the expected equatorial parity) to match modes. This approach could be useful in other problems as well.


Figure~\ref{fig:modes_pg} illustrates the diversity of mode morphologies that emerge in rapidly rotating models. The upper-left and upper-middle panels compare the $m=1$, $\ell_{\mathrm{dom}}=2$ low-frequency mode at $q=0.1$ and $q=0.7$. Its strong horizontal character and interior concentration identify it as a low-order gravito-inertial mode, the continuation of the $\ell=2$, $g_1$ branch, although increasing rotation substantially modifies its geometry. The lower-left and lower-middle panels show the evolution of an axisymmetric radial acoustic mode ($\ell_{\mathrm{dom}}=0$, $m=0$), which we label as part of the $p_2$ sequence. At low rotation rates ($q=0.1$) the eigenfunction is close to the familiar concentric-shell pattern of a classical radial pulsation mode. However, by $q=0.4$, the mode has clearly developed additional angular structure. This indicates rotational mode mixing, whereby the original $\ell=0$ parent couples to higher even-degree components 
allowed by symmetry, most notably $\ell=2$ and $\ell=4$. 
The upper-right panel provides an example of a higher-degree $g$-mode candidate, with $\ell_{\mathrm{dom}}=8$, while the lower-right panel shows
 a high-frequency mode with $\ell_{\mathrm{dom}}=13$ whose surface-skimming
  morphology is typical of a whispering-gallery acoustic mode \citep{Reese2009}. Together, these examples demonstrate that our 2D calculations recover not only the low-degree gravito-acoustic modes, but also the more complex families expected in rapidly rotating stars.

\label{sec:remaining}

\section{Limitations of the present calculation}
\label{sec:limitations}

The present formulation was designed to focus on the oscillations of rotating, hydrodynamic, adiabatic, and polytropic stars. This is a natural stepping stone to more realistic and sophisticated models, but it does have several limitations.

Firstly, our mesh resolution used in these test cases only allows for the low-order acoustic and gravito-inertial modes to be computed with high accuracy. The computation of higher-order modes is more demanding. We can use adaptive mesh-resolution in the FEM for different regions in the stellar interior, and adopting the Cowling approximation (see Appendix A) was also found to be useful for high-order gravity modes.

Secondly, the stellar domain used in our oscillation code terminates on a finite-enthalpy isosurface rather than at the (singular) zero-density surface of the polytrope. However, this boundary condition is physically controlled and can be moved outward systematically, but high-order acoustic and gravity modes can be sensitive to this choice. A robust final error budget should therefore vary $r_{\rm cut}$, and the spatial resolutions and iterative tolerances in both the SCF and eigenvalue calculations. 

Thirdly, the background is barotropic, uniformly rotating, and is described by a single $N=3$ polytropic equation of state.
Real stars can be baroclinic, differentially rotating, 
compositionally stratified, and non-adiabatic effects can be important. These ingredients alter both the equilibrium and the oscillations and would require additional fields or couplings to be introduced.

Finally, mode identification becomes intrinsically 
difficult in a dense spectrum, which is what is typically obtained in rapidly rotating models. 
Spherical degree $\ell$ is no longer an exact quantum number 
at finite rotation, and nearest-frequency tracking can 
fail due to the appearance of avoided crossings. The present continuation method already combines frequency prediction, parity, and the similarity of the angular-energy fingerprint,
 which is sufficient to isolate low-order multiplets through most of the sequence. This is still not equivalent to a direct full-eigenfunction overlap, which would be more robust for continuation. It would also be worthwhile to revisit our approach with finer steps in $q$. 


\section{Future developments and applications} 
\label{sec:prospects}

The principal motivation for using a finite-element (or spectral-element) formulation is that the variational structure is not tied to modelling oblate spheroidal geometries. There are several natural extensions of the present proof-of-concept.

Firstly, a natural extension would be to replace the polytropic 
equilibrium with realistic stellar background models. Because the weak formulation only requires spatial fields such as $\rho$, $P$, $\Gamma_1$, and the effective gravity, the FEM solver can in principle 
be coupled to a two-dimensional rotating stellar model like ESTER \citep{Rieutord2016}. 
Convenient 2D stellar models can also be constructed using the 1D stellar structure code MESA \citep[e.g.][]{Paxton2011,Paxton2013} and the transformation given by the RUBIS code \citep{Hou2023}, which would be an excellent approach for adiabatic oscillation calculations. 

Going beyond 2D, the advantage of FEM makes the application to calculate the 3D oscillation modes of (rotationally and) tidally distorted stars possible. The primary numerical advantage of FEM is that the mesh can conform directly to the Roche potential-like stellar surface without devising a new global coordinate mapping, and it naturally avoids having to explicitly compute the $\ell$- and $m$-couplings that would be required to study such models with a code based on spherical harmonic expansions.
In general, the (rotationally- and) tidally-deformed equilibrium models will not be axisymmetric, so $m$ ceases to be an exact quantum number. In such cases, the present two-dimensional fixed-$m$ model must be replaced by a fully three-dimensional FEM calculation. The equilibrium stellar structure 
of such a synchronised and tidally distorted star can be 
constructed using a similar SCF method \citep{Sirotkin2009,Fellay2024}. Other methods such as the Kippenhahn-Thomas method have also been explored in \citet{Springer2013}. These can be used to construct equilibrium models
for computing the modes using a similar FEM approach to the one we have presented here. 



The FEM framework can also be extended to include tidal forcing of linear oscillation modes, in addition to modelling the tidal distortion of the equilibrium structure. The corresponding inhomogeneous problem can be solved similarly to the method used in \citet{Dewberry2023}, but without the need to work out any generalized coordinate mapping. The same matrices we have computed in this work (Eq. 42) also provide a direct route to studying dynamical tides in uniformly-rotating polytropes. Instead of solving only the homogeneous eigenproblem, one may solve
\begin{equation}
    \left(A_0+\sig A_1-\sig^2 B\right)U=F_{\rm tide},
    \label{eq:forced_tide}
\end{equation}
for each potential component or orbital forcing harmonic, where $F_{\rm tide}$ is the tidal force. This would allow direct calculations of tidally excited waves/oscillations, computation of tidal overlap integrals, and tidal dissipation of oscillation modes in a centrifugally and
tidally distorted star. In eccentric or asynchronous binaries, a three-dimensional equilibrium or time-dependent forcing can be incorporated while retaining the same finite-element infrastructure.

Further extensions could include differential rotation, magnetic fields, and non-adiabatic energy transport. These would introduce additional couplings and, in the magnetic and non-adiabatic cases, additional state variables. Differential rotation would add spatially dependent advection and shear terms, while non-adiabatic or dissipative physics would in general lead to a genuinely non-self-adjoint eigenproblem. Nevertheless, the sparse polynomial/generalized eigenvalue framework used here would remain applicable.

The extension of the weak-form
equation \eqref{mixed} to include these effects is straightforward, and requires considerably less effort than a new derivation
of the strong-form equations and their implementation for such equilibrium models in fully spectral methods. The present calculation is therefore intended not only as a solver for uniformly-rotating polytropes, which provides an independent method to the existing codes used in asteroseismology, but also as a testbed for the development of solution methods for more general multidimensional stellar oscillation problems.

\section{Conclusions}
\label{sec:conclusions}
We have presented a formulation for the computation of linear adiabatic oscillations of a centrifugally distorted, uniformly-rotating polytrope as a two-dimensional (for each azimuthal wavenumber $m$) 
finite-element eigenvalue problem. We have solved the
Eulerian gravitational potential perturbation simultaneously along with a density-weighted displacement as our solution fields. The equilibrium stellar model has been computed using the self-consistent field method, and a conforming two-domain mesh separates the oblate stellar interior from an exterior vacuum extending to a spherical boundary (chosen at $R_{\Phi}=4R_{\rm eq}$), while a common continuous finite-element representation of the perturbed gravitational potential
couples the stellar and vacuum regions naturally.

Our method adopts the density-weighted displacement
$\boldsymbol{u}=\rho^{1/2}\boldsymbol{\xi}$, which is expanded using a real helical representation to describe the solution for each $m$. This preserves the equatorial symmetry of the wave-restoring operator and yields real sparse matrices after finite-element discretisation. The resulting quadratic eigenvalue problem is solved using efficient linear algebraic methods with the PETSc/SLEPc libraries. 

We have presented a hierarchy of tests that have validated our implementation. In the spherical limit, 
our two-dimensional code closely matches the same low-order $g$, $f$, and $p$ modes predicted by 1D stellar oscillation codes. The remaining differences in our calculation
are at the fractional level of $\sim10^{-4}$ for representative modes and are consistent with the finite spatial resolution used in both the oscillation solver and the equilibrium mesh. These differences are somewhat larger than those reported between the spectral codes TOP and ACOR \citep{Ouazzani2012}. This is expected because global spectral methods can converge much more rapidly than a fixed-order finite-element discretization for the smooth solutions and regular geometries considered here. Higher-order finite elements or a spectral-element formulation could substantially improve the accuracy of the mode frequencies, while retaining the geometric flexibility of an element-based method. Another benefit of finite- or spectral-element methods is their natural ability to enable the resolution of localised features in the domain, such as jumps in buoyancy frequency due to chemical/compositional gradients, or when matching stellar radiative and convective layers \citep[e.g.,][]{Guo26}. 

At finite rotation rates, we have compared our oscillation-mode frequencies for an $N=3$ polytrope with $\Gamma_1=5/3$ against predictions from the TOP code \citep{Reese2006a}. For rapidly rotating models, the low-$\ell$ axisymmetric mode branches agree to better than $2.7\times10^{-4}$ over the range $0\le q\le0.6$. This agreement demonstrates that the FEM calculation simultaneously captures the centrifugal distortion, Coriolis dynamics, and perturbed self-gravity in a
strongly non-spherical stellar geometry. Our calculations can also be viewed as an independent verification of the results of \citet{Reese2006a}, since our chosen method is completely different from theirs.

The present work therefore establishes a complete
two-dimensional adiabatic FEM framework for rapidly rotating
polytropes. Unlike approaches tied to a spherical harmonic
representation of the equilibrium geometry, the variational
formulation can be extended naturally to more complicated stellar shapes. This provides a basis for subsequent applications to realistic rotating stellar models, magnetized configurations, and fully three-dimensional tidally distorted stars.

\appendix

\section{Complementary 2D oscillations using the Cowling Approximation}

\begin{figure*}
\centering
\includegraphics[width=0.49\textwidth]{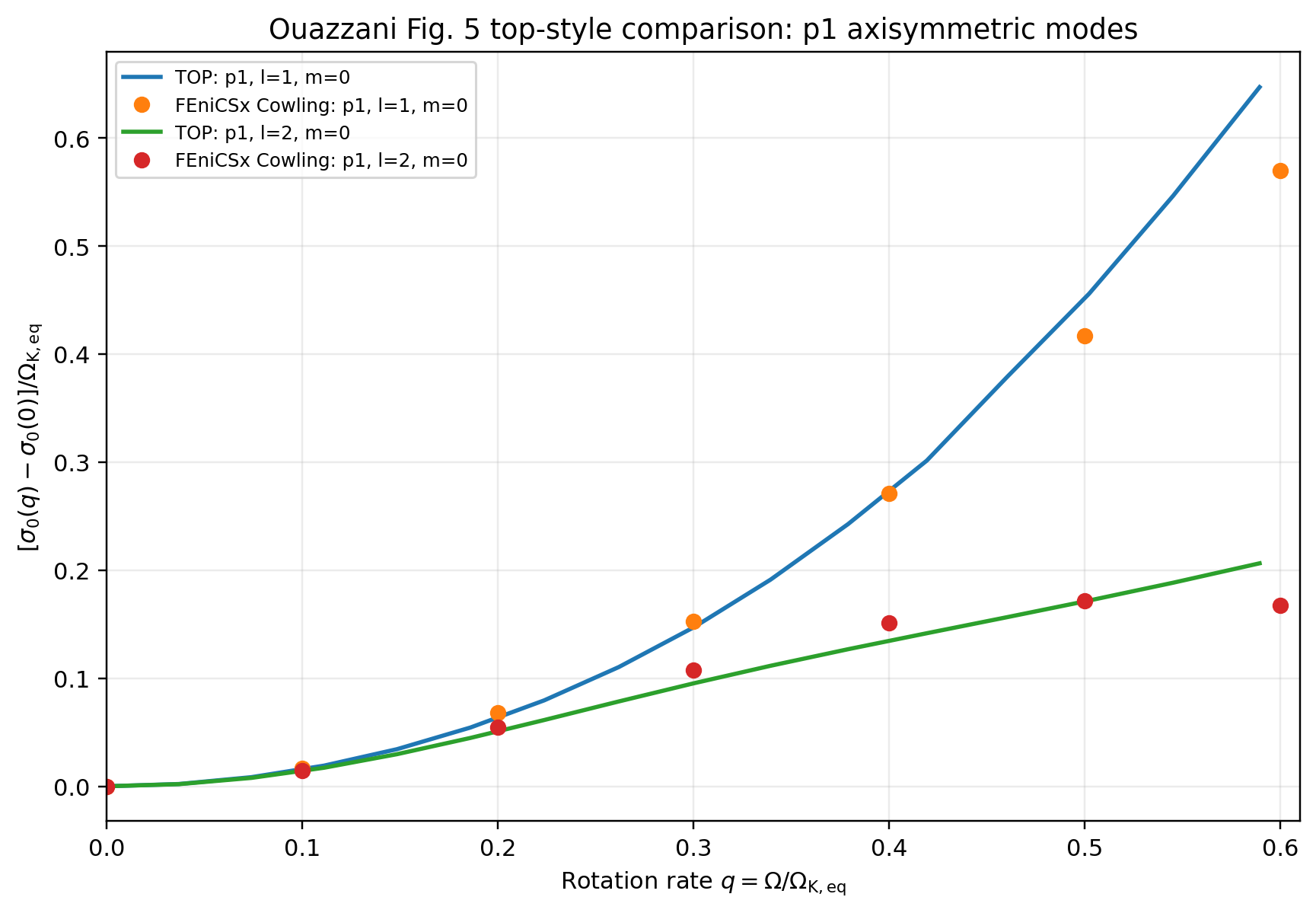}\hfill
\includegraphics[width=0.49\textwidth]{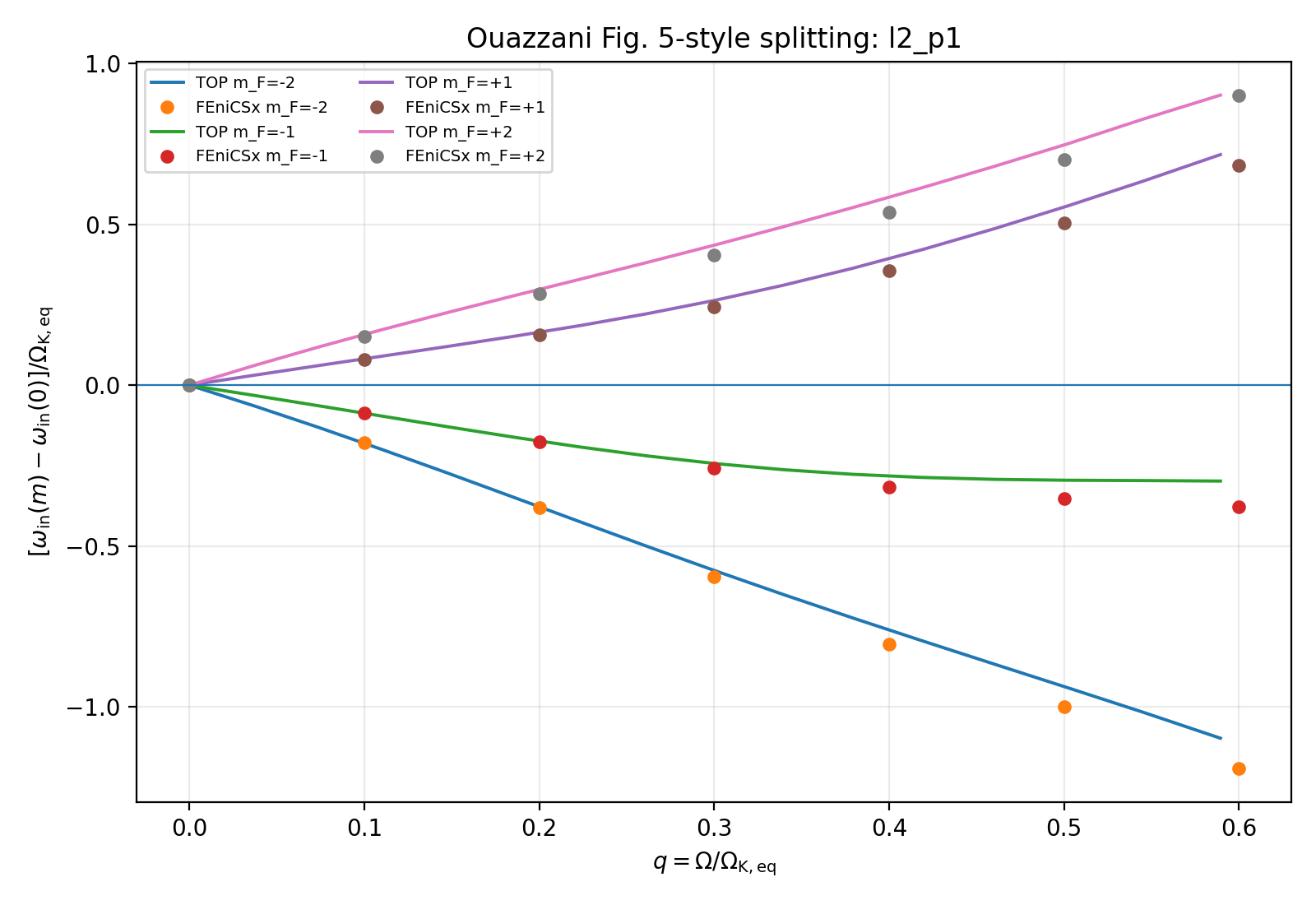}\hfill
\caption{Comparison with the Reese eigenfrequencies in the Cowling approximation for the $p_1$ family. 
Left: the axisymmetric diagnostic, showing the rotational shifts of the
 $\ell=1$ and 2 centroid branches. 
 Right: the complete $\ell=2$ inertial-frame splitting using
  $m_{\rm F}=-2,-1,+1,+2$ in the present FEniCSx convention; 
  each TOP/Reese curve has been evaluated with $m_{\rm TOP}=-m_{\rm F}$. 
  Solid curves are TOP/Reese results and symbols are the Cowling-approximation calculations. 
  Agreement is strongest at low and moderate rotation, while systematic differences increase as the star becomes more distorted.}
\label{fig:fig5_p1_cowling}
\end{figure*}

\subsection{Comparison with Reese/TOP}
\label{subsec:reese}

We also perform calculations of 2D oscillations adopting the Cowling approximation and several complementary tests to validate our results. First, the non-rotating two-dimensional problem is compared with an independent one-dimensional shooting solution and its exact-$m$ degeneracy checked numerically. Second, the absolute non-rotating
frequencies are compared with the results of \citet{Asai2015}, which also adopted the Cowling approximation, and the relative difference is at the level of $0.01\%$. Third, finite-rotation frequency shifts and five-component
$m=-2,\ldots,+2$ sets are compared with the TOP/Reese calculations. Figure A1 shows the comparison of the present FEniCSx Cowling-approximation calculation (dots)
 with the full-gravity Reese/TOP results of \citet{Reese2006a} (lines)
for the $\ell=1$ and 2 $p_1$ branches,
while the right panel shows the complete $m=-2,\ldots,+2$ 
multiplet structure for $\ell=2$. 
The agreement is excellent, although not as close as results from the full-gravity calculation, and a systematic difference remains, especially for moderate and fast rotation.
 
At rapid rotation some Reese/TOP and FEniCSx curves bend 
or encounter mode coupling at different values of 
$q$. This is expected because the Cowling approximation shifts the unperturbed mode spacings, so the location of an 
avoided crossing need not be the same in the two 
calculations. 

\section{Details of the weak formulation and construction of the matrix eigenvalue problem}

We briefly derive the weak formulation used in the finite-element
calculation. We start from the strong-form momentum equation \eqref{eq:strong_form} and the Poisson equation \eqref{eq:poisson_distribution} for $\boldsymbol{u}$ and $\delta \Phi$,
and let $\boldsymbol{v}$ and $\psi$ denote the displacement and gravitational potential test functions, respectively.  Multiplying the momentum equation by
$\boldsymbol{v}^{*}$ and integrating over the stellar volume, the
pressure-gradient term is integrated by parts, such that
\begin{equation}
-\int_{V_\star}\boldsymbol{v}^{*}\cdot\nabla S\,{\rm d}V
=
\int_{V_\star}S\,\nabla\cdot\boldsymbol{v}^{*}\,{\rm d}V
-
\int_{\partial V_\star}S
(\boldsymbol{v}^{*}\cdot\boldsymbol{n})\,{\rm d}A .
\end{equation}
Combining the first volume integral term with
$-\tfrac12 S\boldsymbol{v}^{*}\cdot\nabla\ln\rho$ gives
$S(\boldsymbol{u})Q(\boldsymbol{v})$.  For the barotropic polytropic
equilibrium, $\nabla\ln\rho\parallel\boldsymbol{g}_{\rm eff}$, so that
expansion of $S$, $Q$, and $R$ gives
\begin{equation}
S(\boldsymbol{u})Q(\boldsymbol{v})
+R(\boldsymbol{u})
(\boldsymbol{v}\cdot\boldsymbol{g}_{\rm eff})
=
c_s^2Q(\boldsymbol{u})Q(\boldsymbol{v})
+
(\boldsymbol{u}\cdot\boldsymbol{g}_{\rm eff})
D(\boldsymbol{v})
+
D(\boldsymbol{u})
(\boldsymbol{v}\cdot\boldsymbol{g}_{\rm eff}),
\end{equation}
where $D(\boldsymbol{u})=\nabla\cdot\boldsymbol{u}$.
At the stellar surface, the condition $\Delta P=0$ implies
$Q(\boldsymbol{u})=0$.  Since the boundary is an isenthalpy surface
and $\boldsymbol{g}_{\rm eff}=\nabla H$ is normal to it, the boundary
contribution can be written in symmetric form.

After analytic integration over the azimuthal coordinate, with the
common factor $2\pi$ omitted, the hydrodynamic restoring form is 
\begin{align}
a_{\rm h}(\boldsymbol{v},\boldsymbol{u})
={}&
\int_{\mathcal D_\star}
s\left[
c_s^2Q(\boldsymbol{u})Q(\boldsymbol{v})
+
(\boldsymbol{u}\cdot\boldsymbol{g}_{\rm eff})
D(\boldsymbol{v})
+
D(\boldsymbol{u})
(\boldsymbol{v}\cdot\boldsymbol{g}_{\rm eff})
\right]\,{\rm d}s\,{\rm d}z
\nonumber\\
&+
\int_{\Gamma_\star}
s\,
\frac{
(\boldsymbol{u}\cdot\boldsymbol{g}_{\rm eff})
(\boldsymbol{v}\cdot\boldsymbol{g}_{\rm eff})
}{
|\boldsymbol{g}_{\rm eff}|
}\,{\rm d}\ell .
\end{align}

The perturbed-gravity term in the momentum equation is left without integration by parts and defines the displacement--potential coupling,
\begin{equation}
\begin{aligned}
c(\boldsymbol{v},\phi)
&=
\int_{V_\star}
\rho^{1/2}\boldsymbol{v}^{*}\cdot\nabla\delta\Phi\,{\rm d}V
\\
&=
\int_{\mathcal D_\star}
s\rho^{1/2}
\left[
v_s\partial_s\phi
+v_z\partial_z\phi
+\frac{m}{s}w_v\phi
\right]
\,{\rm d}s\,{\rm d}z .
\end{aligned}
\end{equation}
where $dV=s\,{\rm d}s\,{\rm d}z\,{\rm d}\phi$.
For the Poisson equation, multiplication by $\psi^{*}/(4\pi G)$ and
integration by parts gives
\begin{equation}
c(\boldsymbol{u},\psi)
+k_\Phi(\psi,\phi)=0,
\end{equation}
where
\begin{equation}
k_\Phi(\psi,\phi)
=
\frac{1}{4\pi G}
\int_{\mathcal D_\star\cup\mathcal D_{\rm vac}}
s\left[
\partial_s\psi\,\partial_s\phi
+\partial_z\psi\,\partial_z\phi
+\frac{m^2}{s^2}\psi\phi
\right]
\,{\rm d}s\,{\rm d}z .
\end{equation}
Thus, the displacement--potential coupling in the momentum equation and the corresponding coupling in the Poisson equation are represented by the same bilinear form with the trial and test functions interchanged. 

The Coriolis and mass forms are
\begin{equation}
a_1(\boldsymbol{v},\boldsymbol{u})
=
\int_{V_\star}
\boldsymbol{v}^{*}\cdot
\left(
-2{\rm i}
\boldsymbol{\Omega}\times\boldsymbol{u}
\right)
\,{\rm d}V
=
\Omega\int_{\mathcal D_\star}
s\left(v_+u_+-v_-u_-\right)
\,{\rm d}s\,{\rm d}z ,
\end{equation}
and
\begin{equation}
b(\boldsymbol{v},\boldsymbol{u})
=
\int_{\mathcal D_\star}
s\left(
\frac12v_+u_+
+\frac12v_-u_-
+v_zu_z
\right)
\,{\rm d}s\,{\rm d}z .
\end{equation}
The complete mixed weak problem is therefore to find
$(\sigma,\boldsymbol{u},\phi)$, not identically zero, such that
\begin{equation}
a_{\rm f}(\boldsymbol{v},\boldsymbol{u})
+c(\boldsymbol{v},\phi)
+c(\boldsymbol{u},\psi)
+k_\Phi(\psi,\phi)
+\sigma a_1(\boldsymbol{v},\boldsymbol{u})
-\sigma^2 b(\boldsymbol{v},\boldsymbol{u})
=0
\end{equation}
for all admissible $(\boldsymbol{v},\psi)$.

\section*{Acknowledgments}
This work was supported by STFC grants ST/W000873/1 and UKRI1179. ZG used ChatGPT (OpenAI) to assist with language editing and the refinement of manuscript text and code. All AI-assisted content was reviewed and verified by the authors, who take full responsibility for the final manuscript and the scientific results presented therein.

\section*{Data availability}
The numerical scripts and data products supporting the figures 
will be made available on Zenodo with the published version of this work.


\bibliographystyle{mnras}
\bibliography{2dfem} 








\bsp
\label{lastpage}

\end{document}